\documentclass[12pt,a4paper]{article}

\usepackage{a4wide}
\usepackage{amsmath}
\usepackage{amssymb}
\usepackage{amsfonts}

\DeclareMathOperator{\csch}{csch}
\DeclareMathOperator{\arccsch}{arccsch}
\DeclareMathOperator{\sgn}{sgn}
\usepackage{graphicx}
\usepackage{exscale}
\usepackage{float}
\usepackage{bbm}
\usepackage[numbers,sort&compress]{natbib}

\makeatletter
\@addtoreset{equation}{section}
\makeatother

\title{\vskip-2cm Energy and Angular Momentum Dependent Potentials with Closed Orbits }
\author{M.\ H.\ Al-Hashimi
\footnote{Contact information: M.\ H.\ Al-Hashimi: hashimi@itp.unibe.ch, Bern, Switzerland,
+41 31 534 8948.}}
\begin{document}
\maketitle
\vspace{-1cm}
\begin{abstract} \normalsize
The Bertrand theorem concluded that;  the Kepler potential, and the isotropic harmonic oscillator potential are the only systems under which
all the orbits are closed. It was never stressed enough in the physical or mathematical literature that this is only true when the potentials are independent of the initial conditions of motion, which, as we know, determine the values of the constants of motion $E$ and $L$. In other words, the Bertrand theorem is correct only when $V\equiv V(r)\neq V(r,E,L)$. It has been derived in this work an alternative orbit equation, which is a substitution to the Newton's orbit equation. Through this equation, it was proved that there are infinitely many energy angular momentum dependent potentials $V(r,E,L)$ that lead to closed orbits. The study was done by generalizing the well known substitution $r=1/u$ in Newton's orbit equation to the  substitution $r=1/s(u,E,L)$ in the equation of motion. The new derived equation obtains the same results that can be obtained  from Bertrand theorem. The equation was used to study different orbits with different periodicity like second order linear differential equation periodicity orbits and Weierstrasse periodicity orbits, where interestingly it has been shown that the energy must be discrete so that the orbits can be closed. Furthermore, possible  applications of the alternative orbit equation were discussed, like applications in Bohr-Sommerfeld quantization, and in stellar kinematics.
\end{abstract}

\newpage

\section{Introduction}
The notion of closed orbit is a fundamental concept that shaped the conceptual development of astronomy, and the very first cosmological models created by the ancient world thinkers,
and later developed by medieval thinkers. Take for example the idea of celestial spheres, where all the astronomical objects circulating earth \cite{Grant},
the model based on observing repletion of the position of the celestial objects which takes place after a certain period of time. This essentially a feature of a closed orbit.
The significant development made by  Kepler's three laws \cite{Holton,Gold} cannot be comprehended without the notion of closed orbit explicitly or implicitly.
A well known historical fact, that Newton developed his gravitational theory, in order to find a mathematical discerption to the planetary motion which match,
or explain  Kepler's three laws. Of course, in addition to explain the free fall of objects. The brilliant concept of potential function was invented by Newton
in order to explain the behavior of a particle moving under the action of gravity. Newton found that the successful potential for this purpose is
\begin{equation}
\label{GravPot}
V(r)=-\frac{GmM}{r},
\end{equation}
where $m$ and $M$ are two masses separated by the distant $r$, and $G$ is the gravitation constant, a universal constant that does not depend on the initial conditions,
like energy and angular momentum. This is a property that we have to keep in mind to understand the rest of this work. It is very important here to remember that the choice
of $V(r)$ is totally an empirical choice \cite{Jammer}. It has been made to account for the properties of gravity as they were discovered by all the thinkers prior to Newton.
Their thoughts was driven by observation, or logical reasoning, or by both.

From the Newton's second law, the equation of motion for a single particle with mass $m$ under the action of a arbitrary force $\vec{F}(\vec{r})$ takes the following form
\begin{equation}
\label{NewtSecond}
m\frac{d^{2}\vec{r}}{dt^{2}}=\vec{F}(\vec{r})=-\vec{\nabla} V(\vec{r}),
\end{equation}
By integrating the above equation with respect to time we get the total energy $E$ as a  constant of motion. Then eq.(\ref{NewtSecond}) gives
\begin{equation}
\label{EneExp}
E=\frac{m}{2}\dot{\vec{r}}\cdot \dot{\vec{r}}+ V(\vec{r}).
\end{equation}
For a central force,
\begin{equation}
\label{centeralF}
 V(\vec{r})=V(r), \hspace{10 mm} \vec{F}(\vec{r})=\frac{\vec{r}}{r}F(r),
\end{equation}
as a result of the above equation, we get
\begin{equation}
\label{LConVect}
 \vec{r}\times\vec{F}(\vec{r})=\frac{d}{dt}\left(\vec{r}\times m \vec{v}\right) =\frac{d\vec{L}}{dt}=0,
\end{equation}
which gives the angular momentum vector $\vec{L}$ as another constant of motion. The direction of $\vec{L}$ as well as the magnitude $L$ are constant with respect to time,
and can written in terms of the angular velocity $\dot{\varphi}$ as
\begin{equation}
\label{LCon}
 L =m r^2 \dot{\varphi}.
\end{equation}

In general, solving the equation of motion eq.(\ref{NewtSecond}) for most potentials is not a straightforward process, and even when $r(t)$ can be obtained, the expression is too complicated to be representable. Therefore, a substitution like
\begin{equation}
\label{rSub}
 r(\varphi) =\frac{1}{u(\varphi)},
\end{equation}
would lead to a solvable equation of motion. For a central force, eq.(\ref{NewtSecond}) can be written as \cite{Gold}
\begin{equation}
\label{Norbit}
\frac{d^2u}{d\varphi^2}+u=-\frac{mF(u^{-1})}{L^2u^2}.
\end{equation}
The above equation known as Newton's orbit equation. For the Kepler problem,
\begin{equation}
\label{Fkepler}
 F(r) =-\frac{\kappa}{r^2}=-\kappa u^2,
\end{equation}
the solution of eq.(\ref{Norbit}) is
\begin{equation}
\label{KeplerSol}
 u(\varphi)=\frac{1}{r} =\frac{\kappa m}{L^2}\left(1+e\cos(\varphi-\varphi_0)\right),
\end{equation}
where $e$ is the eccentricity of the elliptical orbit, which is given by the following relation
\begin{equation}
    e=\sqrt{1+\frac{2EL^2}{m\kappa^2}},
\end{equation}
and $\varphi_0$ is the initial value of the angle variable $\varphi$. From now on, we will take $\varphi_0=0$. For the case of the isotropic harmonic oscillator the force is
\begin{equation}
\label{Fkepler}
 F(r) = m \omega^2 r=\frac{m\omega^2}{u},
\end{equation}
the solution of eq.(\ref{Norbit}) is
\begin{equation}
\label{OscSol}
 \frac{1}{r^2} =\frac{E m}{L^2}\left(1+f\cos(2\varphi)\right),
\end{equation}
where
\begin{equation}
    f=\sqrt{1-\frac{\omega^2L^2}{E^2}}.
\end{equation}

The above two examples are particularly important because they lead to a closed orbit. In fact Bertrand's theorem \cite{Bertrand} proves that the Kepler force,
and the isotropic harmonic oscillator force are the only two systems where all the orbits are closed. This is correct regardless of the initial conditions
(which include initial position, and initial velocity that fixed the value of $E$ and $L$). In his theorem, Bertrand only considered forces of the form
\begin{equation}
    F(r)=c r^n,
\end{equation}
where $c$ is a constant that does not depend on constants of motion like $E$ and $L$. In the previous examples, $n=-2$, $c=-\kappa$ for the Kepler system,
and $n=1$, $c=m\omega^2$ for the isotropic harmonic oscillator system. The proof of the theorem based on guarding the stability of a circular orbit that could
be perturbed by a small oscillation. Bertrand have shown that, these two systems are the only ones that do not lead to an orbit that can spiral down to the origin \cite{Gold}.
I have to mention here, that the event of a particle going through the origin, seems to catastrophic to some physicists, especially before the development of the new quantum mechanics,
where a particle can have a zero angular momentum for the $s$-state. However, a particle passing through the center of force is not forbidden mathematically.

Aside from solutions for the Newton's orbit equation that leads to closed obits, there are other studies that aiming at finding potentials for which eq.(\ref{Norbit})
has solutions in terms of known functions. For example, Whittaker have shown that for $n=5,3,0,-4,-5,-7$, then $r(\varphi)$  can be expressed in terms of elliptical functions
\cite{Whittaker} (p83). His work is a result of collecting many studies that go back to the nineteen century. A more generalized approach by Broucke \cite{MAHOMED1} have shown that
 potentials with multiple terms of different powers of $r$ can lead to solution in terms of known functions.
\subsection{angular momentum and Energy dependent Potentials}
Systems with potentials that depends on the initial conditions can be found in many examples in the physics literature. One of the prominent example is the general relativity correction
 to the Newton gravitational potential. Its importance is due to its success in accounting for the perihelion shift for mercury's orbit. According to general relativity, for weak
 gravitational field, slow particle, and an orbit with small eccentricity $e$, the gravitational potential can be approximated to the form \cite{Adler,Jetzer}
\begin{equation}
\label{EinCorr}
    V(r)=-\frac{GMm}{r}-\frac{GML^2}{mr^3c^2}.
\end{equation}
This means that the corrected gravitational potential does depend on the initial conditions of motion through $L$.

Angular momentum dependent potentials are also discussed in the context of quantum field theory, where the scattering amplitude be approximated to a potential using
Born approximation. For example the M{\o}ller scattering has an amplitude correspond to spin-orbit interaction potential between two electrons. This potential depends
on the relative angular momentum between the two electrons explicitly \cite{Sakurai}. This is in addition to countless, other examples where the potential depends on angular momentum,
total energy, or both, which can obtained from quantum field theories. The energy momentum dependent potentials as leading term and not as a mere correction, was also discussed by
when the work of Broucke was further expanded to more potentials using symmetry as tool of investigation \cite{MAHOMED1}.

\subsection{Closed orbits and Bohr-Sommerfeld quantization }
The notion of closed or periodic orbit is central in the development of the old quantum mechanics. In Bohr hydrogen like atom, the electron moves in circular
closed orbits around the nucleus. In Sommerfeld hydrogen like atom, the closed orbits are elliptical. For closed orbit, and as a result of mere observation, the old quantum
 mechanics rules of quantization are
\begin{equation}
\label{Bohr-quan1}
    \oint p_\varphi d\varphi=h n_\varphi, \hspace{10mm} n_\varphi=1,2,3,...
\end{equation}
where $h$ is the Plank constant. Also
\begin{equation}
\label{Bohr-quan2}
    \oint p_r dr=h n_r, \hspace{10mm} n_r=0,1,2,...,
\end{equation}
where $p_\varphi,p_r$ are the conjugate generalized momentums to the generalized coordinates $\varphi$ and  $r$ respectively.  In general, we can write
\begin{equation}
\label{Bohr-quanGeneral}
    \oint p_i dq_i=h n_i,
\end{equation}
where $ n_i=0,1,2,...$ for lateral motion, and  $ n_i=1,2,3,...$  for rotational motion \cite{Schiff,Gold}.
\section{The potential from the periodic equation of motion }
This work is based on deriving the expressions of potentials from the periodicity of the motion of a particle as an input.
To achieve that, instead of the usual substitution that lead to Newton's orbit equation $r=1/u$ eq.(\ref{Norbit}) we use a more general expression
\begin{equation}
\label{ruRelation}
  r(\varphi)=g(u(\varphi),E,L),
\end{equation}
It is important to chose
\begin{equation}
\label{rReg}
    r=g(u,E,L)\geq 0, \hspace{10 mm} g(u,E,L)\in \Re \forall u \in [u_{min},u_{max}].
\end{equation}
Several differential equations with periodic solutions are considered. What we mean by periodic solution in this work is
 \begin{equation}
\label{PeriodicGeneral}
 u(\varphi)= u(\varphi+2n\pi),
\end{equation}
where $n$ a non-zero positive integer. For second order differential equation orbits.
\begin{equation}
\label{PeriodicDiff}
 \frac{d^2u}{d\varphi^2}=\gamma (u,E,L),
\end{equation}
where $\gamma (u,E,L)$ is a function that characterized the kind of periodicity of the solution of the differential equation,
as a function of $u$ and the initial conditions that specify $E$ and $L$. For example, $\gamma (u,E,L)=-\lambda^2 u+a(E,L)$ is corresponds
to a solution of the form $u=a(E,L)\lambda^{-2}+b(E,L)\cos\lambda \varphi$, where $\lambda$ is a rational number. Another example for
$\gamma (u,E,L)=6u^2-g_2(E,L)/2$ that leads to a periodic solution in terms of the Weierstrasse function, which will be studied in more details later.
\\
By using eq.(1.3) and eq.(\ref{ruRelation}), we get
\begin{equation}
\label{u-prime}
 \frac{du}{d\varphi}=\pm\frac{g(u,E,L)}{L g^{\prime}(u,E,L)}\sqrt{2g(u,E,L)^2m(E-V(u,E,L))-L^2},
\end{equation}
by using the chain rule, and substituting for $u^{\prime \prime}(\varphi)$ from eq.(\ref{PeriodicDiff}),$\gamma (u,E,L)$ can be expressed as
\begin{eqnarray}
\label{gamma-eq}
\frac{d^2u}{d\varphi^2}=\frac{du}{d\varphi}\frac{d}{du}\frac{du}{d\varphi}=\gamma (u,E,L),
\end{eqnarray}
The above equation can be integrated with respect to $u$. Further more, by substituting for $u^{\prime}(\varphi)$ from eq.(\ref{u-prime}) then the above equation leads to
following expression for the potential
\begin{eqnarray}
\label{Vu-eq}
V(u,E,L)=E-\frac{L^2}{2m g(u,E,L)^2}-\frac{L^2 g^{\prime}(u,E,L)^2}{m g(u,E,L)^4}\Omega(u,E,L),
\end{eqnarray}
where
\begin{eqnarray}
\label{Omega-eq}
\Omega(u,E,L)=\int_a^u \gamma (\zeta,E,L)d\zeta.
\end{eqnarray}
The function $\Omega(u,E,L)$ is named here as the periodicity characterization function. It represents an input that is restricted by the periodicity of the orbit given by
$\gamma(u,E,L)$. This phrase will be more clear later on in the rest of this article. The number of expressions for $V(u,E,L)$ for a given  $\Omega(u,E,L)$
is infinite, that is because there is infinite possibilities of choosing $g(u,E,L)$. The expression of $V(r,E,L)$ can be retrieved if an inverse solution
\begin{eqnarray}
\label{gInv}
g^{-1}(r,E,L)=u,
\end{eqnarray}
exists. The same result of eq.(\ref{Vu-eq}) can be obtained using different root. By using  eq.(\ref{NewtSecond}) and eq.(\ref{LCon}), for central force,
it can be proved that the force expression $F(r)$ is given by the following equation
\begin{equation}
 \label{rPhi}
 F(r)=\frac{L^2}{m r^4}\frac{d^2r}{d\varphi^2}-\frac{2L^2}{m r^5} \left(\frac{dr}{d\varphi}\right)^2-\frac{L^2}{m r^3}.
\end{equation}
Then using the above equation with eq.(\ref{rReg}) leads to the same expression in eq.(\ref{Vu-eq}).
\\
Another form of eq.(\ref{Vu-eq}) can be obtained by substituting
\begin{equation}
\label{ruRelation2}
  s(u,E,L)=\frac{1}{g(u,E,L)},
\end{equation}
then eq.(\ref{u-prime}) can be written as
\begin{equation}
\label{us-prime}
u'(\varphi)=\sqrt{\frac{2Em-L^2 s(u(\varphi),E,L)^2-2mV(u(\varphi),E,L)}{L^2 s'(u(\varphi),E,L)^2}},
\end{equation}
and $V(u,E,L)$ can be written as
\begin{equation}
\label{Vu2-eq}
V(u,E,L)=E-\frac{L^2}{2m }\left(s^{2\prime}(u,E,L)^2\Omega(u,E,L)+s(u,E,L)^2\right).
\end{equation}
Again here the following condition must hold
\begin{equation}
\label{rReS}
    \frac{1}{r}=s(u,E,L)\geq 0, \hspace{10 mm} s(u,E,L)\in \Re \forall u \in [u_{min},u_{max}].
\end{equation}
The expression given by eq.(\ref{Vu-eq}) is suitable for certain problems, while the expression given by eq.(\ref{Vu2-eq})is more for others, however, both of them will be called alternative orbit equation.
\\
It is important to note at this stage that Bertrand's theorem is applicable only on potentials of the form $V(r,E,L)=V(r)$ which will be called "Bertrand potentials".
\section{Second Order Linear Differential Equation Orbits}
For second order linear differential equations with periodic solution, the eq.(\ref{PeriodicDiff}) has the following form
\begin{equation}
\label{PeriodicDifLinear}
 \frac{d^2u}{d\varphi^2}+u\lambda^2=a(E,L),
\end{equation}
then the solution is
\begin{equation}
 \label{PeriodicDifLinearS}
 u(\varphi)=a(E,L)/\lambda^2+b(E,L)\cos\lambda \varphi.
\end{equation}
From the above equation and eq.(\ref{Omega-eq}) we get
\begin{equation}
\label{OmegaLin}
\Omega(u,E,L)=-\frac{u^2\lambda^2}{2}+a(E,L)u-c(E,L),
\end{equation}
where $c(E,L)$ is an arbitrary constant. Therefore eq.(\ref{Vu2-eq}) for this case can be written as
\begin{equation}
\label{Vu2Cos-eq}
V(u,E,L)=E+\frac{L^2}{2m }\left(2s^{\prime}(u,E,L)^2\left(c(E,L)+\frac{u^2\lambda^2}{2}-a(E,L)u\right)-s(u,E,L)^2\right).
\end{equation}
From eq.(1.3), and by using eq.(\ref{PeriodicDifLinearS}) and eq.(\ref{Vu2Cos-eq}), we get
\begin{equation}
\label{Aexp}
 b(E,L)=\pm\frac{\sqrt{-2 \lambda^2 c(E,L)+ a(E, L)^2}}{\lambda^2}.
\end{equation}
As eccentricity is only associated with conic section orbits, we define a more suitable parameter as a generalize conic section
$\tilde{e}(E,L)$ which can be expressed from eq.(\ref{PeriodicDifLinearS}). It is defined by the following relation
\begin{equation}
\label{ecc}
   \tilde{e}(E,L)=\frac{\lambda^2 b(E,L)}{a(E,L)}
\end{equation}

There are countless cases for $V(u,E,L)$, let us first consider the Bertrand potentials case
\begin{equation}
\label{VBertrand}
  V(u,E,L)=V(u)=V(r),
\end{equation}
where the expression of the potential is independent of the constants of motion $E$ an $L$. A solution for this case is
\begin{equation}
\label{vucase}
    s(u,E,L)=s(u)=u^n,
\end{equation}
by substituting for $s(u)$ from eq.(\ref{vucase})into eq.(\ref{Vu2Cos-eq}), the result is
\begin{equation}
\label{vucaseN}
  V(u)=E+\frac{L^2}{2m}\left(n^2\lambda^2-1\right)u^{2n}+\frac{n^2L^2}{m}\left(u^{2n-2}c(E,L)-u^{2n-1}a(E,L)\right),
\end{equation}
The above equation can only be satisfied if $\lambda=1/n$. Moreover, the third or the fourth terms can cancel $E$ only
if $n=1$ or $2n=1/2$. For $n=1$, then $\lambda=1$ and the Kepler potential $V(r)=-\kappa/r$ case is retrieved with
\begin{equation}
\label{Kep}
 s(u)=u=\frac{1}{r}, \hspace{10 mm} a(E,L)=-\frac{Em}{L^2}, \hspace{10mm} c(E, L)= \frac{m\kappa}{L^2},
\end{equation}
the value of $b(E,L)$ can be calculated from eq.(\ref{Aexp}), which gives
\begin{equation}
\label{KepA}
 b(E,L)=\frac{m\kappa}{L^2}\sqrt{1+\frac{2EL^2}{m\kappa^2}},
\end{equation}
which is the same expression that can be obtained from eq.(\ref{KeplerSol}). For $n=1/2$, then $\lambda=2$ and the isotropic harmonic oscillator potential
$V(r)=m\omega^2r^2/2$ case is retrieved, with
\begin{eqnarray}
\label{IHO}
 s(u)&=&\sqrt{u}=\frac{1}{\sqrt{r}}, \hspace{10mm} c(E,L)=\frac{2m\omega^2}{L^2}, \hspace{10mm} a(E, L)= \frac{4Em}{L^2},\hspace{10mm}\nonumber
 \\ b(E,L)&=&\frac{4Em}{L^2}\sqrt{1-\frac{L^2\omega^2}{E^2}}
\end{eqnarray}
which is in agreement with eq.(\ref{OscSol}). As we see here, the results comply with  Bertrand's theorem, as there are no other potentials of the form
$V(u)=V(r)$ except the Kepler and the isotropic harmonic oscillator potentials. However, it must be kept in mind that this treatment only applicable on second order
linear differential equation orbits.

The other class of potentials of the form $V(u,E,L)=V(u,E)$ which shall be named as rescaling potentials. The following substation can lead to such potentials
\begin{equation}
\label{Lind}
 s(u,E,L)=\frac{1}{L}\widetilde{s}(u,E), \hspace{10 mm} c(E,L)=c(E),\hspace{10 mm} a(E,L)=a(E),
\end{equation}
then eq.(\ref{Vu2Cos-eq}) can be written as
\begin{equation}
\label{VuE-eq}
V(u,E)=E+\frac{1}{2m }\left(\widetilde{s}^{2\prime}(u,E)^2\left(c(E)+\frac{u\lambda^2}{2}-a(E)u\right)-\widetilde{s}(u,E)^2\right),
\end{equation}
where $c(E)$ and $a(E)$ are arbitrary functions of the total energy. It must be mentioned here that for such substitution
\begin{equation}
\label{VER}
    V(u,E)=V(rL^{-1},E).
\end{equation}

The simplest non-trivial example for this case is a potential $V(u)=V(rL^{-1})$. A suitable choice for $s(u,E,L)$ for this case is
\begin{equation}
\label{VER}
  s(u,E,L)=\frac{\alpha_0+\alpha_1 u}{L}=\frac{1}{L}\widetilde{s}(u)=\frac{1}{r},
\end{equation}
where $\alpha_0$ and $\alpha_1$ are constants that are independent of $E$ and $L$. In order that the first term in eq.(\ref{VuE-eq}) canceled, $c(E)$ must be chosen as
\begin{equation}
\label{bER}
  c(E)=\frac{\alpha_0^2-2Em}{2\alpha_1^2}.
\end{equation}
For an elegant expression of $V(rL^{-1})$ all the terms of power  a $V(u,E)\sim u^2$, then
\begin{equation}
\label{cER}
  a(E)=-\frac{\alpha_0}{\alpha_1},
\end{equation}
which leads to
\begin{equation}
\label{AER}
  b(E)=\frac{1}{\lambda^2}\sqrt{\frac{\alpha_0^2+\left(2Em-\alpha_0^2\right)\lambda^2}{\alpha_1^2}},
\end{equation}
which leads to
\begin{equation}
\label{rEx1}
  r=\frac{L\lambda^2}{\alpha_0(\lambda^2-1)\sgn \alpha_1+\sqrt{2Em\lambda^2-(\lambda^2-1)\alpha_0^2}\cos\lambda\varphi},
\end{equation}
Accordingly the potential in eq.(\ref{VuE-eq}) has the following expression
\begin{equation}
\label{VEU1}
    V(u)=\frac{u^2 \alpha_1^2}{2m}\left(\lambda^2 -1\right),
\end{equation}
and therefore
\begin{equation}
\label{VER1}
    V(r,E,L)=\left(\lambda^2 -1\right)\left(\frac{\alpha_0^2}{2m}+\frac{L^2}{2mr^2 }-\frac{\alpha_0 L}{2m r}\right).
\end{equation}
It is obvious that for $\lambda=1$ leads to the free particle case. Otherwise, the first term is just a constant shift in energy, the second term
is a centrifugal potential up to a constant $\lambda^2 -1$ , and the third term is a Kepler-like potential with a coupling constant
$-\alpha_0 L/(2m )\sim L$, which is an essential difference from the usual Kepler potential. For real and positive $r$ (see eq.(\ref{rReS})), and using  eq.(\ref{rEx1}),
the energy values for this system is restricted by the following condition
\begin{equation}
\label{ECon1}
 \frac{\alpha_0^2}{2m}\left(\lambda^2 -1\right) \geq E \geq \frac{\alpha_0^2}{2m\lambda^2}\left(\lambda^2 -1\right),
\end{equation}
\vspace{10mm}
\begin{figure}[tbh]
\begin{center}
\includegraphics[bb=360 28 400 400,scale=0.6]{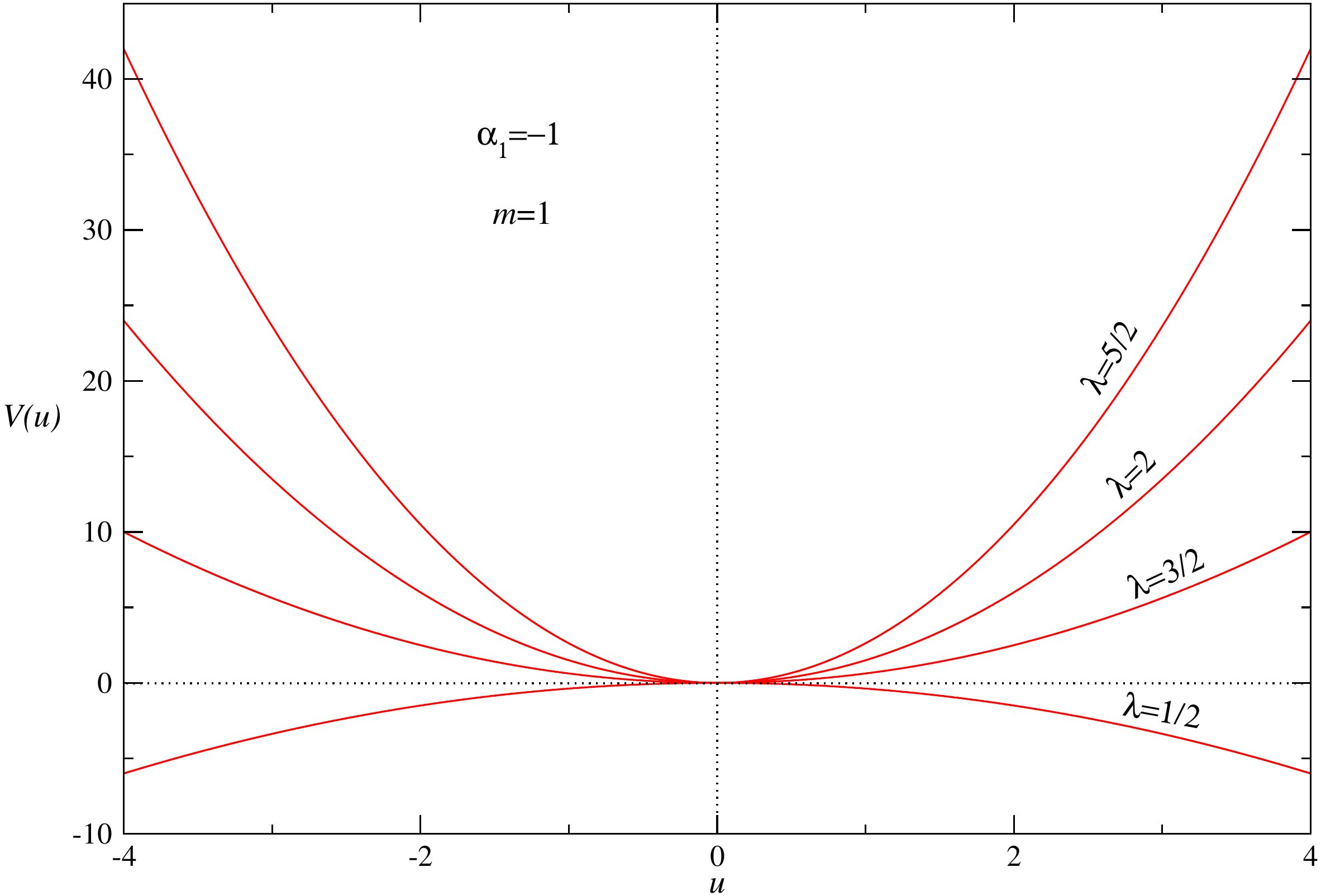}
\vspace{5mm}
\caption{\it The potential $V(u)$ versus $u$ for different values of $\lambda=1/2,3/2,2,5/2$. In all the graphs, $m=1$, and $\alpha_1=-1$.}
\label{fVu1}
\end{center}
\end{figure}
\begin{figure}[tbh]
\begin{center}
\includegraphics[bb=300 28 400 400,scale=0.6]{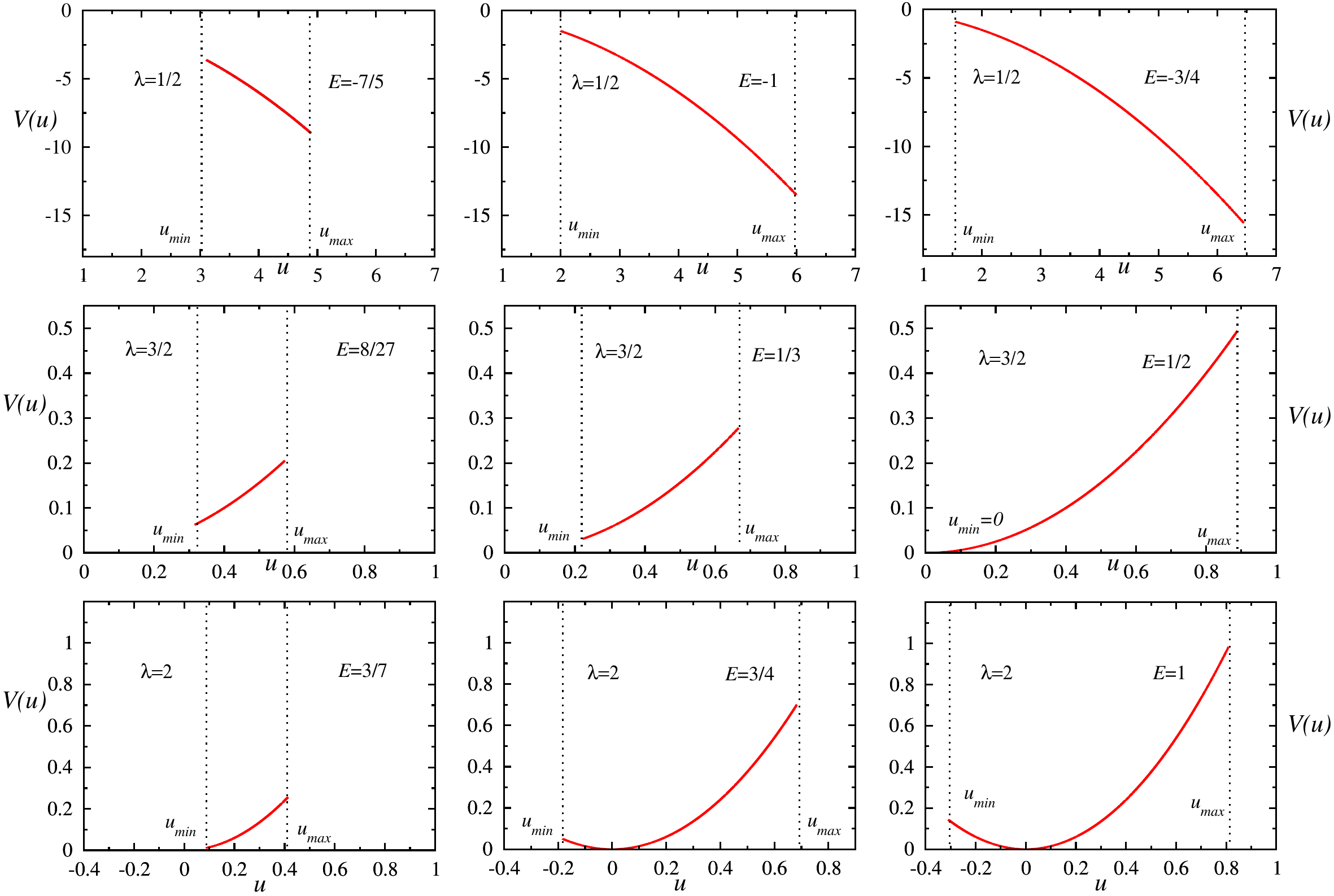}
\vspace{5mm}
\caption{\it The potential $V(u)$ versus $u$ for different values of $\lambda$, and different values of the energy $E$. The values of $E$ are chosen for each value of
$\lambda=1/2,3/2,2$ such that the condition in eq.(\ref{ECon1}) is respected. The variable $u$ has a minimum value of $u_{min}$, and maximum value $u_{min}$,
which are determined by eq.(\ref{PeriodicDifLinearS}), with $\alpha_0=-1$ for $\lambda=1/2$, and $\alpha_0=1$ for $\lambda=3/2,2$. In all the graphs $m=1$, and $\alpha_1=-1$}
\label{fVu2}
\end{center}
\end{figure}
\begin{figure}[tbh]
\begin{center}
\includegraphics[bb=280 28 400 400,scale=0.6]{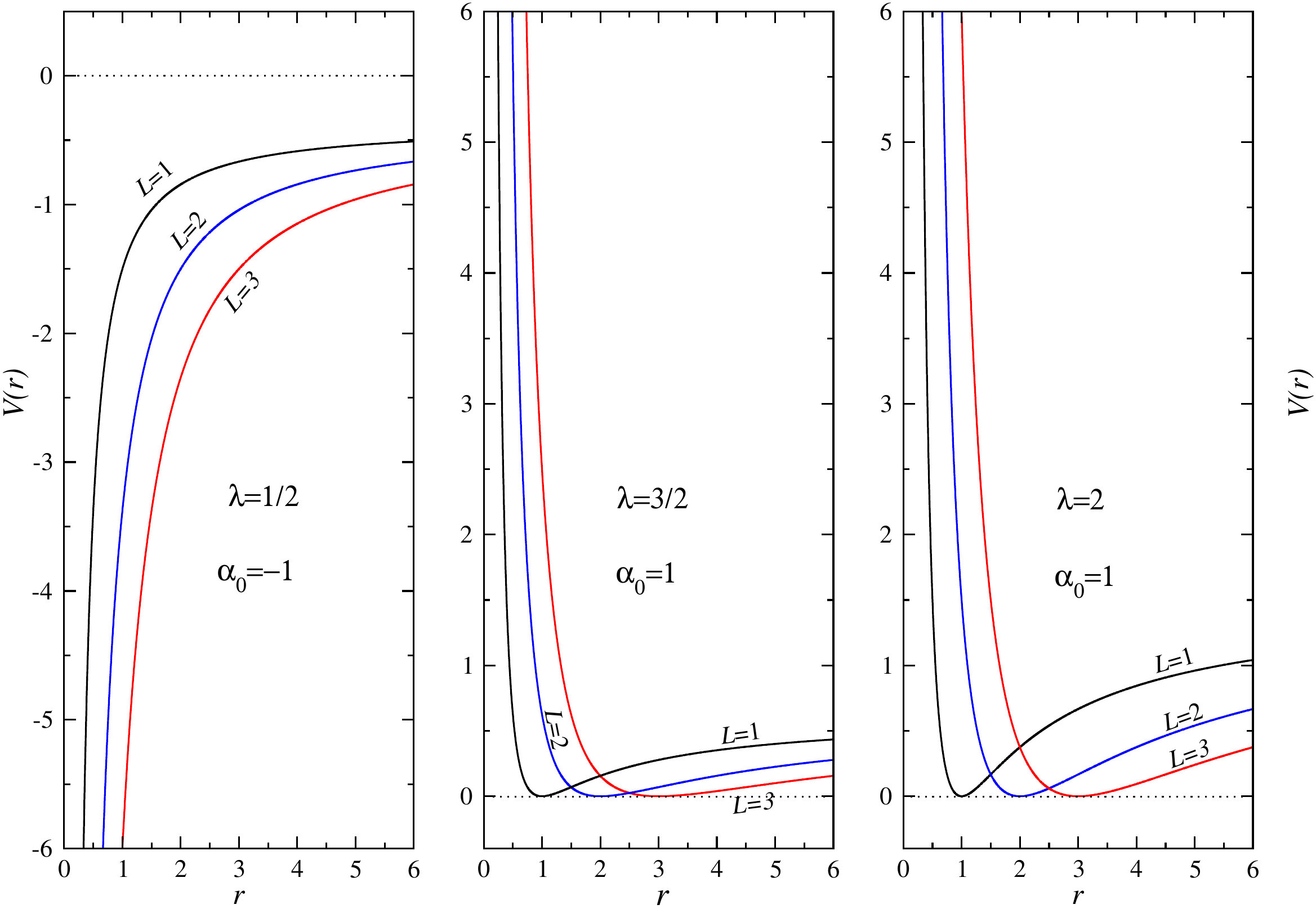}
\vspace{5mm}
\caption{\it The rescaling potential $V(rL^{-1})$ versus $r$ for different values of $\lambda $, and  $L=1,2,3$ for each value of $\lambda=1/2,3/2,2 $.
The value $\alpha_0=-1$  for $\lambda =1/2$, and  $\alpha_0=1$  for $\lambda =3/2,2$. In all the graphs, $m=1$.}
\label{fVr}
\end{center}
\end{figure}
\\
moreover, $\alpha_0<0$ for $\lambda=1/2$, and $\alpha_0>0$ for $\lambda>1$. In figure \ref{fVu1}, the potential $V(u)$ is plotted against $u$ for different values for
$\lambda$ using eq.(\ref{VEU1}).
All the curves are symmetric for any value of $\lambda$. In figure \ref{fVu2}, the potential $V(u)$ is plotted against $u$ for certain values of energy $E$, and different values of
$\lambda$. The energies subject to condition eq.(\ref{ECon1}). In this case $u\in [u_{min},u_{max}]$, the value of $u_{min}$ and $u_{max}$ can be calculated for the given $\lambda$
and $E$ using eq.(\ref{PeriodicDifLinearS}). In figure \ref{fVr}, the potential $V(rL^{-1})$ is plotted against $r$. The curves has been plotted for different value of $\lambda$, and for
$L=1,2,3$ \footnote{It is must be stressed here that choosing the values of $L=1,2,3$ in different graphs has nothing to do with any quantization. It is just a choice for the sake
of simplicity. The same goes for any value of energy $E$, which has been calculated according to our simple choice of $m=1$, and $\alpha_0=\pm 1$.}.
In the left panel it is obvious that $V(rL^{-1})\rightarrow-\infty$ as $r\rightarrow 0$ for $\lambda=1/2$, while in the middle and right panels $V(rL^{-1})\rightarrow\infty$ as
$r\rightarrow 0$ for $\lambda>1/2=3/2,2$. For $\lambda=1/2$, the potential constant $\alpha_0=-1$, which is  in compliance with the condition in eq.(\ref{rReg}),
that demands in this case that $\alpha_0<0$. On the other hand, choosing $\alpha_0>0$ satisfies this condition for $\lambda>3/2,2$.
\begin{figure}[tbh]
\begin{center}
\includegraphics[bb=270 28 400 400,scale=0.6]{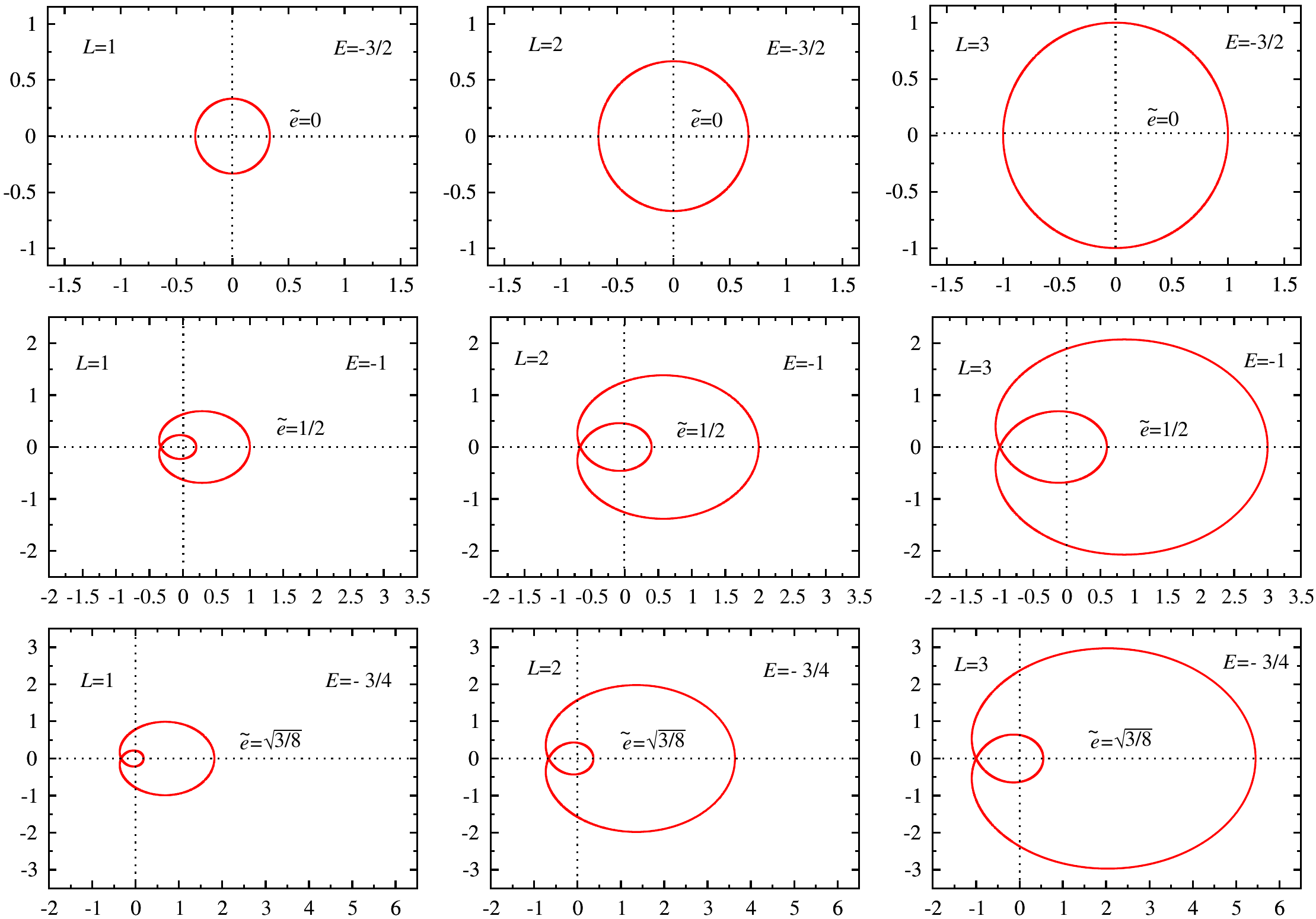}
\vspace{5mm}
\caption{\it Closed orbits for a particle moving under the action of the rescaling potential $V(rL^{-1})$ for $\lambda=1/2$,
and for different values of the energy $E =-3/2,-1,-3/4$. For each value of $E$  there is a graph with $L=1,2,3$ .
In all the graphs, $m=1$, $\alpha_0=-1$, and $\alpha_1< 0$.}
\label{fOrb05}
\end{center}
\end{figure}
\begin{figure}[tbh]
\begin{center}
\includegraphics[bb=300 28 400 400,scale=0.6]{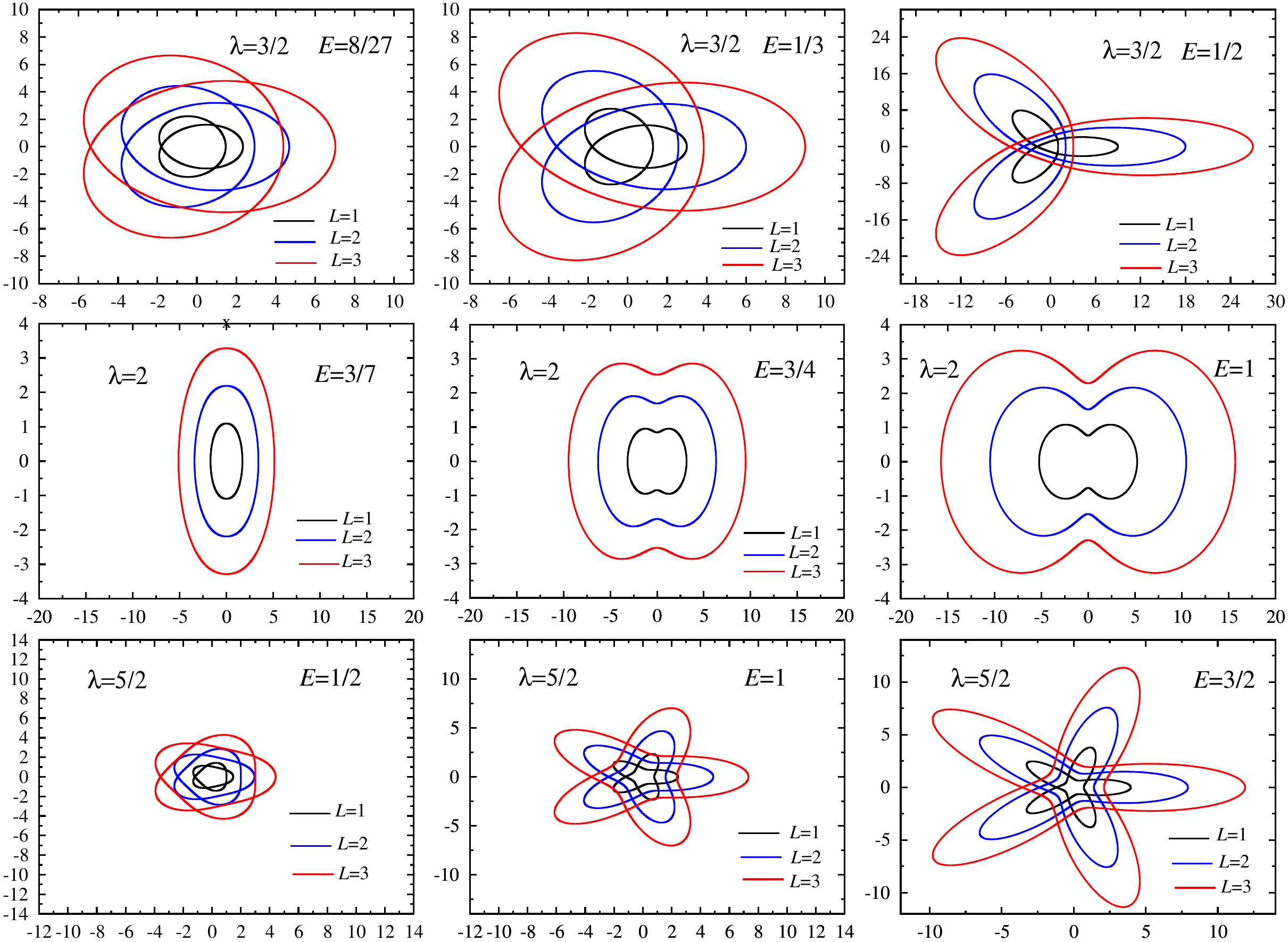}
\vspace{5mm}
\caption{\it Closed orbits for a particle moving under the action of the rescaling potential $V(rL^{-1})$ for different values of  $\lambda$,
and different values of  $L=1,2,3$ correspond to each value of $\lambda =1/2,3/2,2$. In this figure, the energies has been chosen to highlight
the stretching of the orbit with the increase of energy. The calculations were done with, $\alpha_0=-1$ for $\lambda =1/2$, and  with $\alpha_0=1$  for
$\lambda =3/2,2$. For all the graphs  $m=1$,  and $\alpha_1< 0$.}
\label{fOrbAll}
\end{center}
\end{figure}
\begin{figure}[tbh]
\begin{center}
\includegraphics[bb=280 12 400 400,scale=0.6]{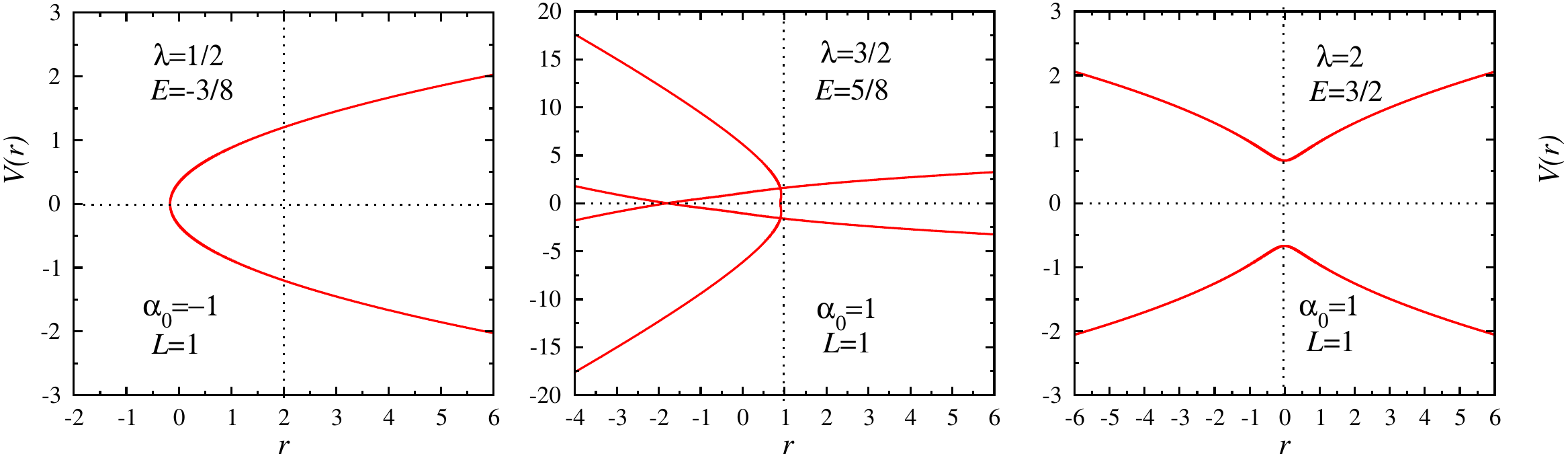}
\vspace{5mm}
\caption{\it Orbits with $r\rightarrow \infty$ at certain point or points for a particle moving under the action of the rescaling potential $V(rL^{-1})$.
The three graphs are for values of $\lambda =1/2,3/2,2$. The value $\alpha_0=-1$  for $\lambda =1/2$, and $\alpha_0=1$  for $\lambda =3/2,2$.
The energies are for the upper limit values, which are $E=-3/8,5/8,3/2$ correspond to $\lambda =1/2,3/2,2$ respectively. In all the graphs  $m=1$, $\alpha_1< 0$, and $L=1$.}
\label{fOrbOpen}
\end{center}
\end{figure}
The last three figures are important for studying the potential for this case. Figure \ref{fOrb05}
is for the orbit with $\lambda=1/2$. For $m=1$, $\alpha_0=-1$, the energy $E=-3/2$ has the lower possible value according to eq.(\ref{ECon1}). Then the orbit is a circle,
with an increasing radius as $L=1,2,3$ increases from the left upper panel to right upper panel. This highlight the motivation of naming $V(rL^{-1})$ as the rescaling potential,
where, for example, the orbit for certain energy and $L=2$ is a copy of the orbit with $L=1$, except it is bigger in size. The other orbits in other graphs of figure \ref{fOrb05}
are for other values of energy $E=-1,-3/4$, where it calculated such that the condition in eq.(\ref{ECon1}) is fulfilled. For these two case, the generalized eccentricity
$\tilde{e}(E,L)=0,1/2,$ and $\sqrt{3/8}$ respectively, which can be calculated from eq.(\ref{ecc}). It is clear that for $\tilde{e}(E,L)\neq 0$, the orbit intersects once with itself. In figure \ref{fOrbAll}, the graphs are for different values of $\lambda$, and different energies. For each value of $\lambda$ and $E$, three different orbits are plotted for
$L=1,2,3$. We note here that for a half integer $\lambda$, the orbit intersect with itself, as one can see that from  figure \ref{fOrb05} while for $\lambda$ an
integer, the orbit does not intersect with it self, as one can see in figure \ref{fOrbAll} for the $\lambda=2$. Figure 6 shows that the lowest values of $E$ leads
to $r\rightarrow \infty$ for $\lambda\varphi=0,\pi, 2\pi,... $. The figures from 3 to 6 show that eq.(\ref{Vu2-eq}) actually works.

Another example of a rescaling potential $V(u,E )=V(r\sqrt{-mE} L^{-1} )$ is a potential that leads to a rescaling factor $L/\sqrt{-mE}$. It can be generated by using the
following substitution
\begin{equation}
\label{sinh1}
\widetilde{s}(u)=A \sinh(\sigma u)=\frac{L}{r}
\end{equation}
in eq.(\ref{VuE-eq}), where $\sigma $ is a constant, which is independent of $E$. To cancel $E$ in the expression of $V(u,E )$, we make the following substitution
\begin{equation}
\label{ConSinh1}
A=\sqrt{-2mE} \hspace{10mm} c(E)=\frac{1}{2\sigma^2 },
\end{equation}
which means that for this case $E<0$. Accordingly, the expression of $V(u,E)$ can be written as
\begin{equation}
\label{VuSinh1}
V(u,E)=Eu\sigma^2\left(2 a(E)-u\lambda^2\right)\cosh(\sigma u)^2.
\end{equation}
It is obvious from the above equation, that the expression of $a(E)$ cannot be manipulated to reduce or further simplify the expression of $V(u,E)$.
On the other hand, it can be chosen such that the expression of $u(\varphi)$ is simpler. Moreover, for a certain expression of $a(E)$, there can be an
upper limit on the value of $\lambda$. For example, for
\begin{equation}
\label{aSinh1}
a(E)=a=\frac{\mu}{\sigma},\hspace{10 mm}\mu=\frac{n}{2}, \hspace{10 mm} n=2,3,4,...,
\end{equation}
which leads to
\begin{equation}
\label{bSinh1}
b(E)=\frac{\sqrt{\mu^2-\lambda^2}}{\lambda^2 \sigma},
\end{equation}
where $\mu > \lambda$. Therefore we get
\begin{equation}
\label{uSinh1}
u(\varphi)=\frac{1}{\lambda^2}\left(\mu+\sqrt{\mu^2-\lambda^2}\cos(\lambda\varphi)\right),
\end{equation}
and
\begin{equation}
\label{rSinh1}
r=\frac{L}{\sqrt{-2mE}}\csch\left[\frac{1}{\lambda^2}\left( \mu+\sqrt{\mu^2-\lambda^2}\right)\cos(\lambda\varphi)\right].
\end{equation}
It is obvious from eq.(\ref{rSinh1}) that the rescaling factor for the orbit is $L/\sqrt{-2mE}$. The expression of $V(rL^{-1},E)$ for this case is
\begin{eqnarray}
\label{VrSinh1}
V(rL^{-1},E)&=&\frac{L^2-2Emr^2}{2mr^2}\arccsch\left(\frac{r\sqrt{-2mE}}{L}\right)\nonumber\\&\times& \left(\lambda^2\arccsch\left(\frac{r\sqrt{-2mE}}{L}\right)-2\mu \right).
\end{eqnarray}
\begin{figure}[tbh]
\begin{center}
\includegraphics[bb=280 28 400 400,scale=0.6]{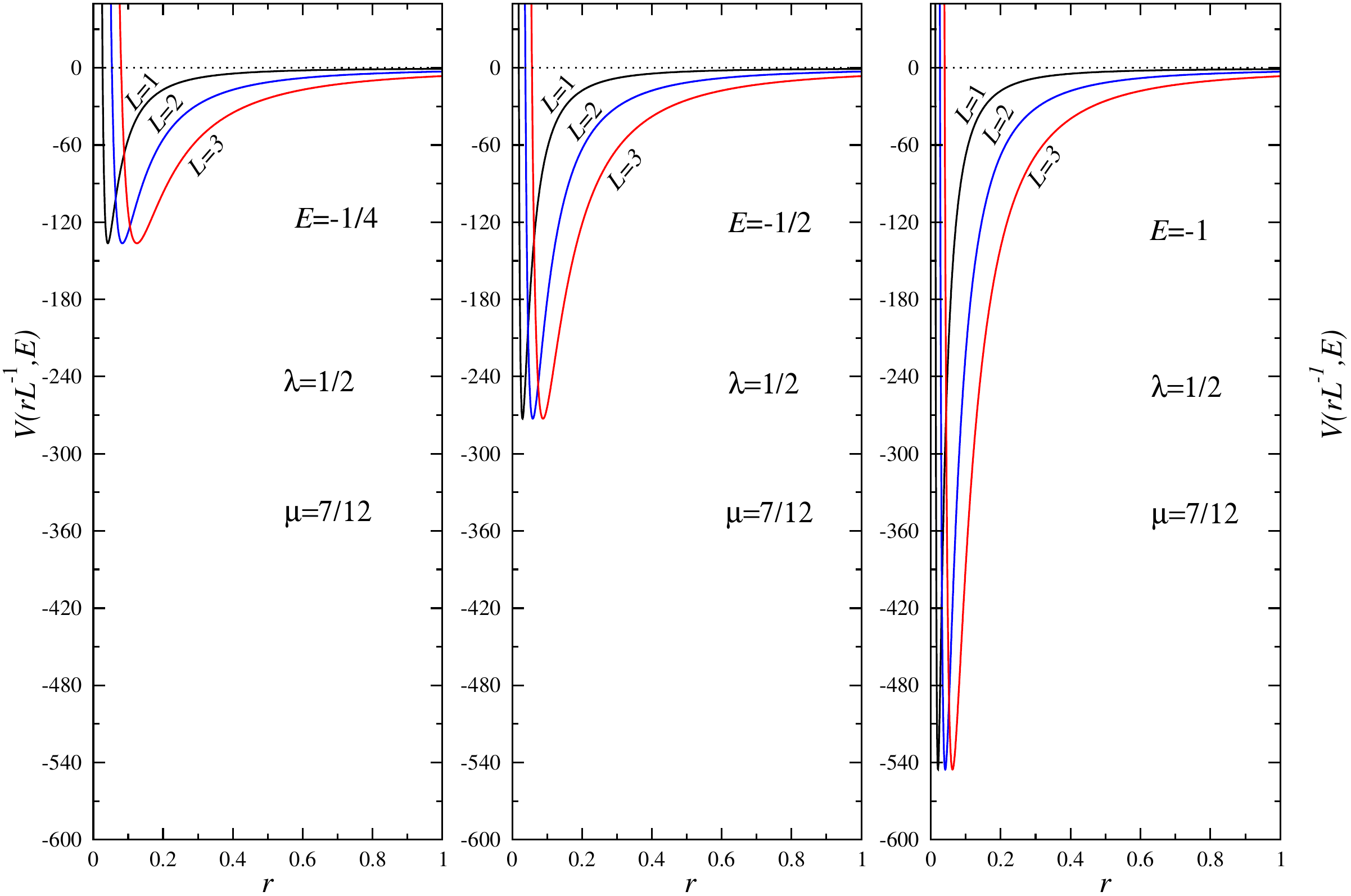}
\vspace{5mm}
\caption{\it The rescaling potential $V(rL^{-1},E)$ versus $r$ for $\lambda =1/2$. Each graph is for different energy $E=-1/4,-1/2,-1$. The curves in each graph are for $L=1,2,3$.
In all the graphs $m=1$,  $\mu=7/12$, and the value of the generalized eccentricity $\tilde{e}(E,L)=\sqrt{3}/2$.}
\label{fOrbVrSinh}
\end{center}
\end{figure}
In figure \ref{fOrbVrSinh}, the potential $V(rL^{-1},E)$ is plotted against $r$ by using eq.(\ref{VrSinh1}).
The curves has been plotted for one value of  $\lambda=1/2$. Roughly speaking, all the graphs show that the energy is more attractive as
$|-E|$ is getting bigger. Again here, it must be kept in mind the units in all the figures were ignored.
\begin{figure}[tbh]
\begin{center}
\includegraphics[bb=260 28 400 400,scale=0.6]{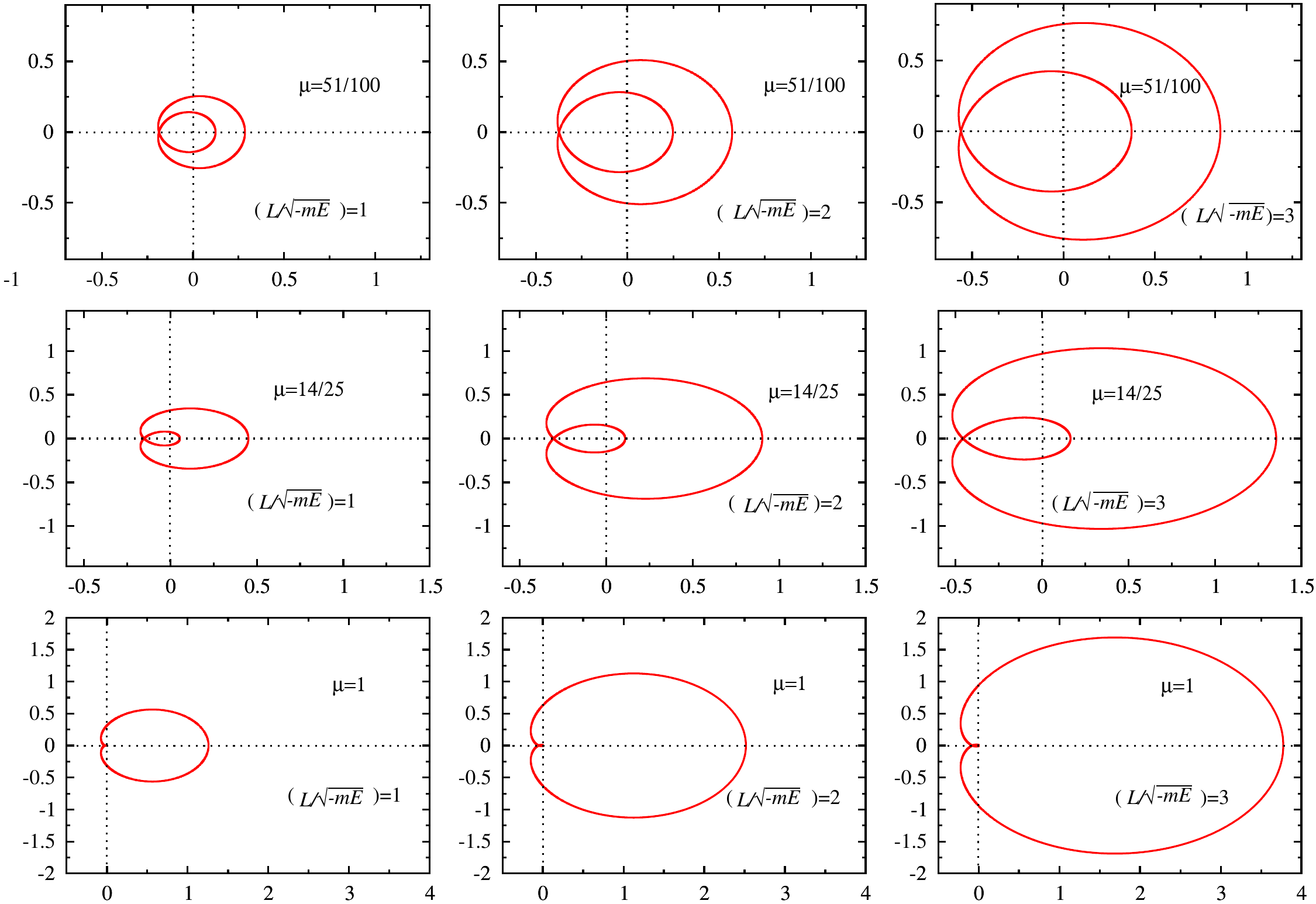}
\vspace{5mm}
\caption{\it Closed orbits for a particle moving under the action of the rescaling potential $V(rL^{-1},E)$ for $\lambda=1/2$.
The figure has three rows of graphs, for the first row $\sigma=1/32$, for the second row  $\sigma=1/8$, and for the third $\sigma=1$.
In each row, there are three orbits. The rescaling factor for each orbit are $L/\sqrt{-mE}=1,2,3$ respectively. For all the graphs $m=1$.}
\label{fOrbSinhEq1}
\end{center}
\end{figure}

\begin{figure}[tbh]
\begin{center}
\includegraphics[bb=240 28 400 400,scale=0.6]{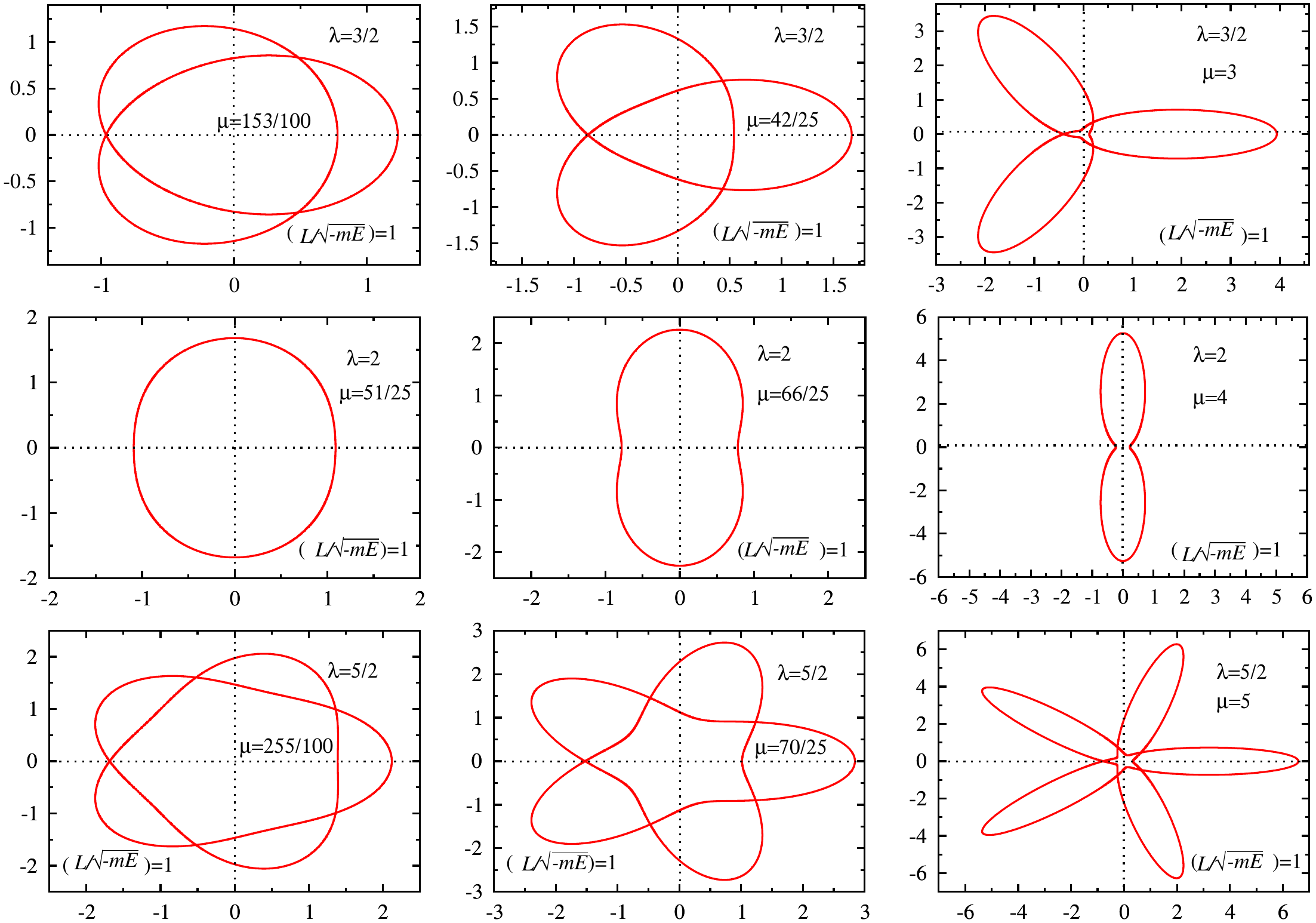}
\vspace{5mm}
\caption{\it  Closed orbits for a particle moving under the action of the rescaling potential $V(rL^{-1},E)$.
The figure has three rows of graphs, for the first row $\lambda=3/2$, for the second row  $\lambda=2$, and for the third $\lambda=5/2$.
In each row there are three orbits, for $\mu=153/100,42/25,3$ respectively. For all the graphs $m=1$, and the rescaling factor $L/\sqrt{-mE}=1$ .}
\label{fOrbSinhEq2}
\end{center}
\end{figure}
Figures \ref{fOrbSinhEq1},\ref{fOrbSinhEq2} show again that eq.(\ref{Vu2-eq}) leads to closed orbits. The orbits in figure \ref{fOrbSinhEq1} are for $\lambda=1/2$.
In general they have a common shape with with the orbits in figure \ref{fOrb05} (the second and the third rows), for example, in both cases the orbit
intersect once with itself. On the other hand, they are not similar, especially with larger values of $\mu$. This is simplify because eq.(\ref{rEx1})
is different than eq.(\ref{rSinh1}). Moreover, for the $\sinh$-potential, the rescaling factor can be taken as $L/\sqrt{-mE}$ instead of $L$, as one can
readily see from  eq.(\ref{rSinh1}). For this case, for any $E$, and any $L$, we can obtain the same orbit, as long as $L/\sqrt{-mE}$ is fixed.
In physical terms, for a given $L/\sqrt{-mE}$, the increasing of $L$ that leads to an increase of the centrifugal force can be balanced by an increase of
$|-E|$ which increase the attraction force (see figure \ref{fOrbVrSinh}), such that the orbit is unchanged. This is true as long as $L/\sqrt{-mE}$ unchanged.

In figure \ref{fOrbSinhEq2}, the orbits are for different values of $\lambda=3/2,2,5/2$. For each value of $\lambda$, three graphs have been plotted for $\mu=153/100,42/25$ and $\mu=3$.
In all the graphs $L/\sqrt{-mE}=1$. It is clear that orbits stretched as $\mu$ increases for fixed rescaling factor $L/\sqrt{-mE}=1$.
\subsection{The potential of the form $V(r,E,L)=\sum_{n=1}^4\frac{A_n(E,L)}{r^n}$}
If the condition in eq.(\ref{Lind}) is dropped, then the expression of is $s(u,E,L)$ is not necessarily separable. Let us assume that $s(u,E,L)$ has the following expression
\begin{equation}
\label{sVEL}
s(u,E,L)=\frac{u}{\alpha_0(E,L)+\alpha_1(E,L) u}=\frac{1}{r}
\end{equation}
The energy $E$ in the expression of $V(u,E,L)$ can be canceled out in eq.(\ref{Vu2Cos-eq}) by using the following substitution
\begin{equation}
\label{cVEL}
c(E,L)=-\frac{m E \alpha_0(E,L)^2}{L^2},
\end{equation}
then $V(r,E,L)$ can be written as
\begin{equation}
\label{Vr1VEL}
V(r,E,L)=\sum_{n=1}^4\frac{A_n(E,L)}{r^n},
\end{equation}
where,
\begin{eqnarray}
\label{An}
A_1(E,L)=4E\alpha_1(E,L)-\frac{L^2a(E,L)}{m\alpha_0(E,L)}, \nonumber \\ A_2(E,L)=\frac{L^2}{2m}(\lambda^2-1)-6E\alpha_1(E,L)^2+\frac{3L^2a(E,L)\alpha_1(E,L)}{m\alpha_0(E,L)},
 \nonumber  \\ A_3(E,L)=4E\alpha_1(E,L)^3-\frac{L^2\lambda^2\alpha_1(E,L)}{m}-\frac{3L^2\alpha_1(E,L)^2a(E,L)}{m \alpha_0(E,L)},
 \nonumber  \\ A_4(E,L)=-E\alpha_1(E,L)^4+\frac{L^2\lambda^2\alpha_1(E,L)^2}{2m}+\frac{L^2\alpha_1(E,L)^3a(E,L)}{m \alpha_0(E,L)}.
\end{eqnarray}
There are many special cases for the potential in eq.(\ref{Vr1VEL}), an important case is the case with $\alpha_1(E,L)=0$ and $\lambda=1$, then
\begin{equation}
\label{Vr2VEL}
V(r,E,L)=-\frac{L^2a(E,L)}{m r\alpha_0(E,L) }=-\frac{\kappa(E,L)}{r},
\end{equation}
where
\begin{equation}
\label{kVEL}
\kappa(E,L)=\frac{L^2a(E,L)}{m \alpha_0(E,L) },
\end{equation}
is a totally arbitrary function of $E$ and $L$. The orbit equation for this case can be obtained by calculating $b(E,L)$ from eq.(\ref{Aexp}). The result is
\begin{equation}
\label{KeplerSol}
 r=\frac{\alpha_0(E,L)}{u} =\frac{L^2}{\kappa(E,L) m\left(1+e\cos(\varphi)\right)},\hspace{10mm}   e=\sqrt{1+\frac{2EL^2}{m\kappa(E,L)^2}}.
\end{equation}
When $\kappa(E,L)=\kappa$, then the Kepler potential is retrieved. The different possibilities of choosing $\kappa(E,L)$ lead to important
applications, which will be discussed later in this article.

Another special case can arise when $\alpha_1(E,L)\neq 0$, and $A_3(E,L)=A_4(E,L)=0$, which can be fulfilled when
\begin{equation}
\label{aVEL}
a(E,L)=\mp\lambda\alpha_0(E,L)\frac{\sqrt{-2Em}}{L}, \hspace{10mm} \alpha_1(E,L)=\pm\lambda\frac{L}{\sqrt{-2Em}}.
\end{equation}
Accordingly, the potential function for this case is
\begin{equation}
\label{Vr3VEL}
V(r,E,L)=\mp\frac{\lambda L}{r}\sqrt{\frac{-2E}{m}}+(\lambda^2-1)\frac{L^2}{2mr^2}, \hspace{10mm} \alpha_1(E,L)=\pm\lambda\frac{L}{\sqrt{-2Em}}.
\end{equation}
From eq.(\ref{Aexp}), the value of $b(E,L)=0$. However, this does not mean that the orbit is a circle. From eq.(\ref{aVEL}) it is clear that
$a(E,L)=0$, which makes the system non- physical because $r=0$ for any $\varphi$, thus, $V(r,E,L)$ singular for this case for any value of $\varphi$.
This is an important reminder that not all the choices of $s(u,E,L)$ can lead to a physical solution. Other solutions are dismissed because they
lead to negative or imaginary $r$ for any value of $\lambda$, which means that the condition in eq.(\ref{rReS}) is violated. These cases are
$A_2(E,L)=A_3(E,L)=0$, and $A_1(E,L)=A_2(E,L)=0$.

\begin{figure}[tbh]
\begin{center}
\includegraphics[bb=280 28 400 400,scale=0.6]{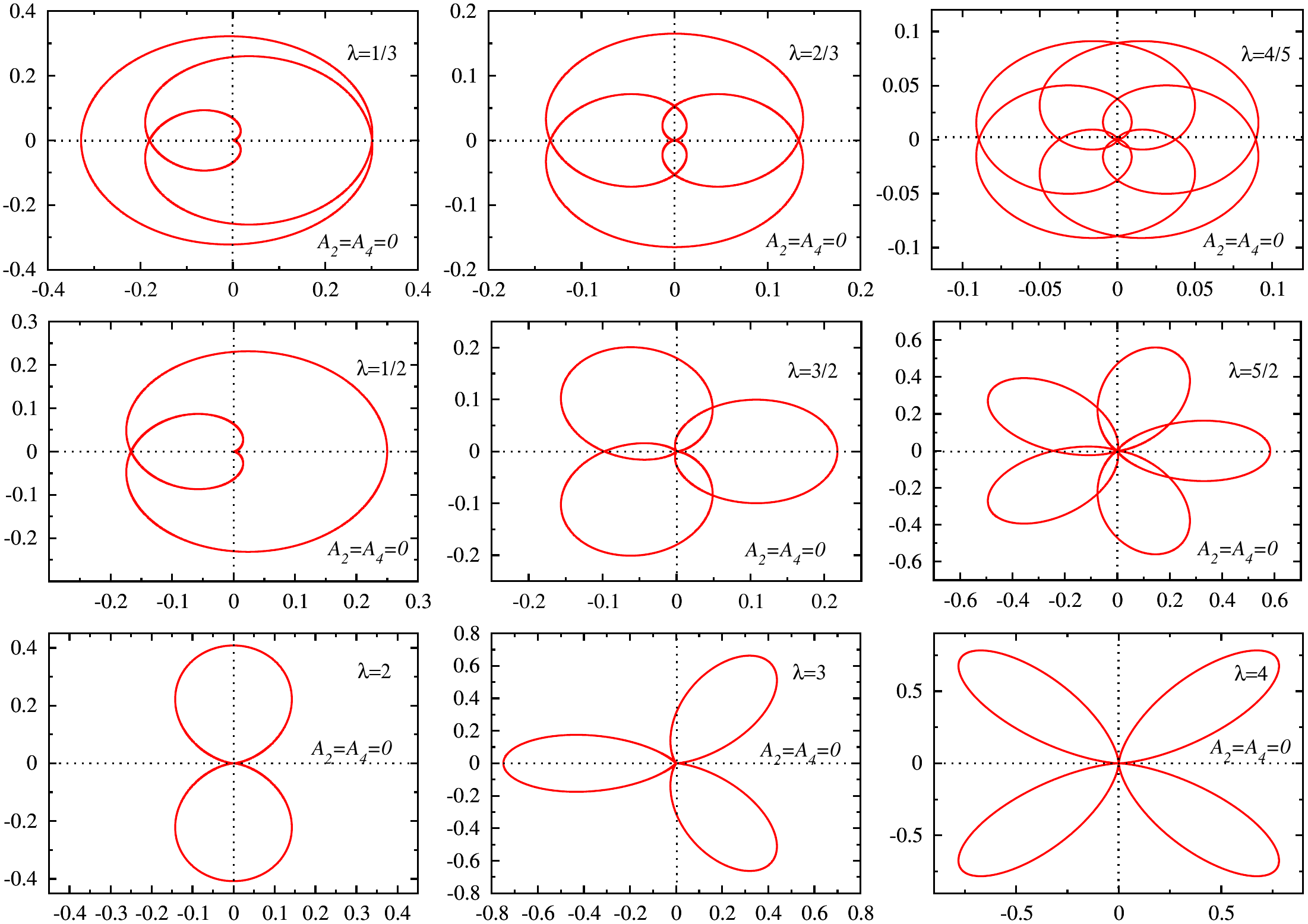}
\vspace{5mm}
\caption{\it Closed orbits for a particle moving under the action of the potential $V(rL^{-1},E)$ for the first physical case with $A_2=A_4=0$
for different values of $\lambda$. For all the graphs $m=1$, $L/\sqrt{-mE}=1$, and $\alpha_0(E,L)>0$.}
\label{fOrbEmpFirst}
\end{center}
\end{figure}

The are three physical cases, \textbf{the first physical cases} is for $A_2(E,L)=A_4(E,L)=0$, with $\alpha_1(E,L),\alpha_0(E,L)>0$. Then, in addition to eq.(\ref{cVEL}),
the following relations are valid
\begin{equation}
\label{a2VEL}
a(E,L)=\mp \frac{\alpha_0(E,L)(1+5\lambda^2)}{L}\sqrt{\frac{-Em}{6(1+2\lambda^2)}}, \hspace{10mm} \alpha_1(E,L)=\pm\frac{L\sqrt{1+2\lambda^2}}{\sqrt{-6Em}}.
\end{equation}
The potential for this case is

\begin{equation}
\label{Vr4VEL}
V(rL^{-1},E)=\mp\frac{L(1+\lambda^2)}{r}\sqrt{\frac{-3E}{2m(1+2\lambda^2)}}\pm\frac{L^3\left( 2\lambda^4-\lambda^2-1\right)}{6mr^3\sqrt{-6Em(1+2\lambda^2)}}.
\end{equation}
The trajectory for this case can be obtained from the following relation for $\lambda<1$
\begin{equation}
\label{rVEL}
r=L\sqrt{\frac{2\lambda^2+1}{-6Em}}\frac{\left(1-\lambda^2\right)(1+\sgn\alpha_0\cos(\lambda\varphi))}{5\lambda^2+1+\sgn\alpha_0(1-\lambda^2)\cos(\lambda\varphi)}
\hspace{10mm} \alpha_1(E,L)=\frac{L\sqrt{2\lambda^2+1}}{\sqrt{-6Em}},
\end{equation}
and the following relation for $\lambda>1$
\begin{equation}
\label{rbVEL}
r=L\sqrt{\frac{2\lambda^2+1}{-6Em}}\frac{\left(\lambda^2-1\right)(1-\sgn\alpha_0\cos(\lambda\varphi))}{5\lambda^2+1+(\lambda^2-1)\sgn\alpha_0\cos(\lambda\varphi)}\hspace{10mm}
\alpha_1(E,L)=-\frac{L\sqrt{2\lambda^2+1}}{\sqrt{-6Em}}.
\end{equation}
The orbits for this case are closed for any $E<0$, as it shown in figure \ref{fOrbEmpFirst}.

\begin{figure}[tbh]
\begin{center}
\includegraphics[bb=280 28 400 400,scale=0.6]{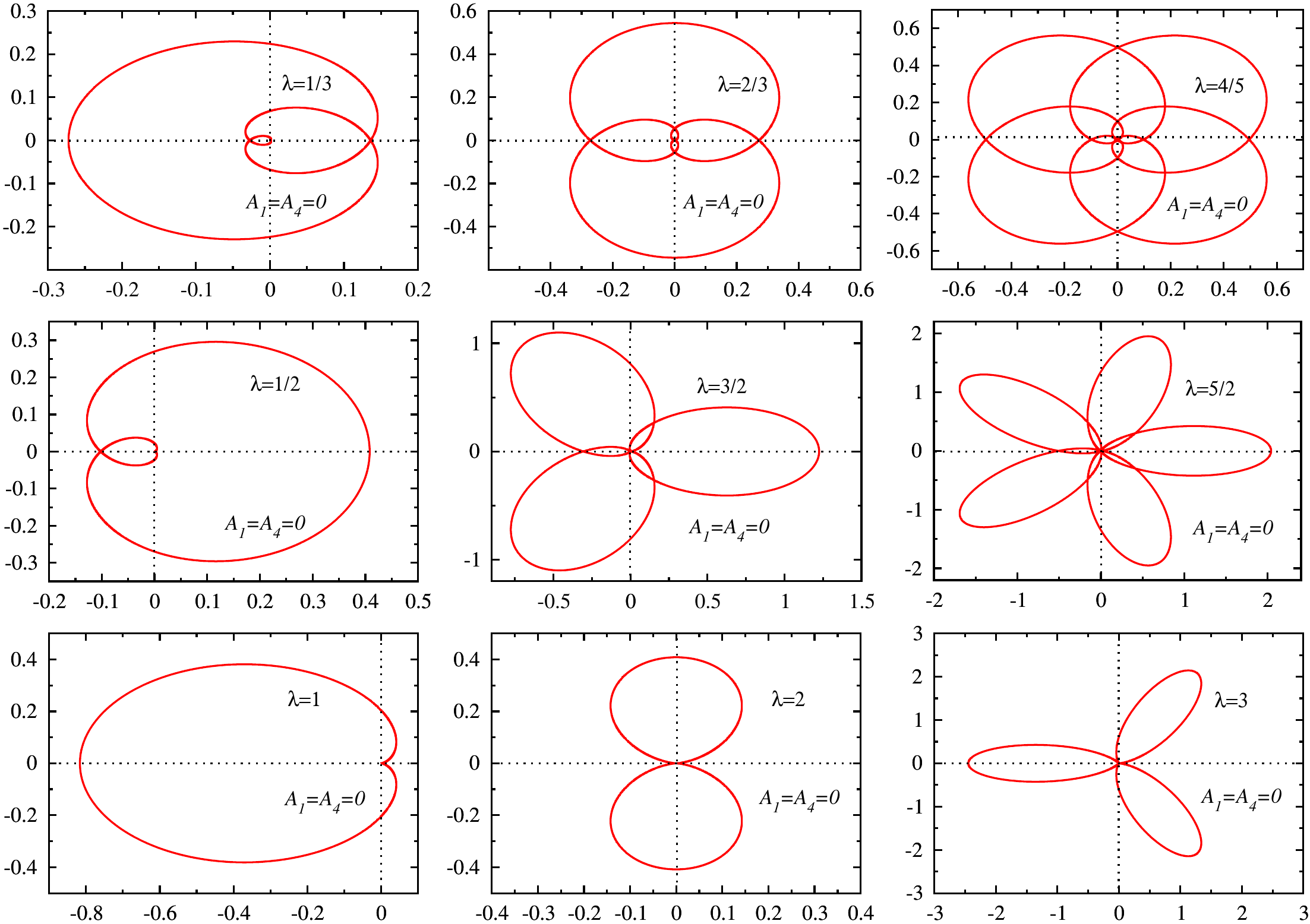}
\vspace{5mm}
\caption{\it  Closed orbits for a particle moving under the action of the potential $V(rL^{-1},E)$
for the second physical case with $A_1=A_4=0$ for different values of $\lambda$. For all the graphs $m=1$, $L/\sqrt{-mE}=1$, and $\alpha_0(E,L)>0$.}
\label{fOrbEmpSecond}
\end{center}
\end{figure}

\textbf{The second physical cases} is for $A_1(E,L)=A_4(E,L)=0$. Then, in addition to eq.(\ref{cVEL}), the following relations are valid
\begin{equation}
\label{a3VEL}
a(E,L)=\mp\sqrt{\frac{-8Em}{3}}\frac{\lambda\alpha_0(E,L)}{L}, \hspace{10mm} \alpha_1(E,L)=\pm\frac{L\lambda}{\sqrt{-6Em}}.
\end{equation}

\begin{equation}
\label{Vr5VEL}
V(rL^{-1},E)=-\frac{L^2(1+\lambda^2)}{2mr^2}\pm\frac{L^3\lambda^3}{3mr^3\sqrt{-6Em}}.
\end{equation}

\begin{equation}
\label{rVEL}
r=\frac{L\lambda}{\sqrt{-6Em}}\frac{\sgn\alpha_0(E,L)-\cos(\lambda\varphi)}{2\sgn\alpha_0(E,L)+\cos(\lambda\varphi)},\hspace{10mm}  \alpha_1(E,L)=-\frac{L\lambda}{\sqrt{-6Em}}.
\end{equation}
The orbits for this case are closed for any $E<0$, as it shown in figure \ref{fOrbEmpSecond}. It is important to note here that the solution with
$\alpha_1(E,L)=L\lambda/\sqrt{-6Em}$ leads to the non-physical case with $r<0$.

\textbf{The third physical case} and the physical case is for $A_1(E,L)=A_3(E,L)=0$, the solution is
\begin{equation}
\label{Vr6VEL}
V(rL^{-1},E)=-\frac{L^2}{2mr^2}(1+\frac{\lambda^2}{2})-\frac{L^4\lambda^4}{64Em^2r^4}, \hspace{10mm} \alpha_1(E,L)=\pm\frac{L\lambda}{2\sqrt{-2Em}}.
\end{equation}
For this case $b(E,L)=0$, then the only closed orbit is a circle with a radius
\begin{equation}
\label{rVEL}
r=\frac{L\lambda}{2\sqrt{-2Em}},
\end{equation}
and,
\begin{equation}
\label{alphaCon}
\alpha_1(E,L)=-\frac{L\lambda}{2\sqrt{-2Em}},
\end{equation}
which means that $r=contant>0$ for any $E<0$ and $L$. As for $\alpha_1(E,L)=L\lambda/{2\sqrt{-2Em}}$, then $r=contant<0$, therefore this solution is discarded.
\vspace{10mm}
\vspace{15mm}
\section{The Weierstrasse Function Periodicity Orbits}
We start this chapter by explaining some basic properties of the Weierstrasse function. There are several
 Weierstrasse functions that share one common feature of being doubly periodic functions, or quasi-periodic. In relation to this work, studying all
 Weierstrasse functions would be a very lengthy process. Therefore we will only study special cases of the Weierstrasse's elliptic function $\wp$
 which is defined by the following relation\cite{Lawden, Akhiezer, Hancock}
\begin{equation}
\label{WeierDef}
    \wp(z;g_2,g_3)=\frac{1}{z^2}+\sum_{m,n=-\infty}^{\infty'}\frac{1}{(z-2m\omega_1-2n\omega_3)^2}-\frac{1}{(2m\omega_1+2n\omega_3)^2},
\end{equation}
where the prime means that terms in the sum give zero denominators are excluded. The half periods in the above equation are $\omega_1$ and $\omega_3$.
It is clear from eq.(\ref{WeierDef}) that
\begin{equation}
\label{WeierPeriod}
    \wp(z+2m\omega_1+2n\omega_3;g_2,g_3)=\wp(z;g_2,g_3),\hspace{10mm} m,n=1,2,3,...
\end{equation}
The Weierstrasse invariants $g_2$ and $g_3$ can be evaluated using the following relations \cite{Wolfram}
\begin{equation}
\label{Weierg}
   g_2=60\sum_{m,n=-\infty}^{\infty'}\frac{1}{(2m\omega_1+2n\omega_3)^4}\hspace{10mm}g_3=140\sum_{m,n=-\infty}^{\infty'}\frac{1}{(2m\omega_1+2n\omega_3)^6}.
\end{equation}
It must be noted here that here that notation used here for different variables are used by most references. The Weierstrasse function $\wp$
satisfies the following two differential equations
\begin{equation}
\label{WeierDiff1}
 \wp''(z;g_2,g_3)-6 \wp(z;g_2,g_3)^2+\frac{g_2}{2}=0,
\end{equation}
and
\begin{equation}
\label{WeierDiff2}
 \wp'(z;g_2,g_3)^{2}-4 \wp(z;g_2,g_3)^3+g_2\wp(z;g_2,g_3)+g_3=0.
\end{equation}
Another feature of $\wp$ that can be obtained from eq.(\ref{WeierDef}) is \cite{Lawden}
\begin{equation}
\label{WeierHo}
 \wp(zt;g_2,g_3)=\frac{1}{t^2}\wp(zt;g_2t^4,g_3t^6).
\end{equation}
From what has been mentioned, it is possible that $\wp$ to be imaginary and negative, what we are interested in, is a positive real $\wp$. Such an example is
\begin{equation}
\label{WeierEx1}
 \wp(z;3,1)=\frac{3}{2}\cot\left(\sqrt{\frac{3}{2}}z\right)^2+1.
\end{equation}
Then periodicity orbit function  $u(\varphi)$ can be written as
\begin{equation}
\label{Weieru}
 u(\varphi)=\wp(\varphi;g_2,g_3),
\end{equation}
Therefore, the only $\wp$ that leads to closed orbits is the one with the periodicity
\begin{equation}
\label{WeierSpecial}
\wp(\varphi;g_2,g_3)=\wp(\varphi+2n\pi;g_2,g_3),\hspace{10mm}n=1,2,3,...,
\end{equation}
which restricts further the values of $g_2$ and $g_3$. For this case, using the above first differential equation eq.(\ref{WeierDiff1}), and eq.(\ref{gamma-eq}) give
\begin{equation}
\label{Weiergamma}
 \gamma(u,E,L)=6 u^2-\frac{g_2(E,L)}{2},
\end{equation}
and therefore eq.(\ref{Omega-eq}) gives
\begin{equation}
\label{WeierOmega}
 \Omega(u,E,L)=2 u^3-\frac{g_2(E,L)u}{2}-c(E,L).
\end{equation}
Accordingly,  eq.(\ref{Vu2-eq}) for this case can be written as
\begin{equation}
\label{WeierVu2-eq}
V(u,E,L)=E+\frac{L^2}{2m }\left(s^{\prime}(u,E,L)^2\left(c(E,L)+2 u^3-\frac{g_2(E,L)u}{2}\right)-s(u,E,L)^2\right).
\end{equation}
Furthermore, using eq.(\ref{us-prime}), eq.(\ref{WeierVu2-eq}) and the second differential equation eq.(\ref{WeierDiff2}) we get
\begin{equation}
\label{Weiercg}
 c(E,L) =\frac{g_3(E,L)}{2}.
\end{equation}
\subsection{The Bertrand theorem and Weierstrasse periodicity}
For this periodicity, to investigate existence of Bertrand potentials such that  $V(u,E,L)=V(u)=V(r)$, it has to be proved that  the expression of the potential is
independent of the constants of motion $E$ an $L$. An ansatz like the one of eq.(\ref{vucase}) can be examined. Then eq.(\ref{WeierVu2-eq}) gives
\begin{equation}
\label{WeiervucaseN}
  V(u,E,L)=E+\frac{L^2}{2m}\left(-u^{2n}+g_3n^2u^{2n-2}+g_2n^2u^{2n-1}-4n^2u^{2n+1}\right).
\end{equation}
For certain choice of $n$, it is possible to cancel $E$ in the above expression. On the other hand, there are no choices for $n$, $g_2$ or $g_3$ that cancel the dependency of the
potential on $L$. Therefore, we can say; for the Weierstrasse periodicity, there is no possibility of constructing a Bertrand potential using an ansatz $s(u,E,L)=s(u)=u^n$.
It is not clear if other choice of $s(u,E,L)$ will lead to a  Bertrand potential. Many choices has been examined and failed. Therefore, at this point, we can say that the
second order differential equation periodicity is the only periodicity that leads to a  Bertrand potential.
\subsection{A simple example for the Weierstrasse periodicity orbits}
The simplest example for this case is
\begin{equation}
\label{WeierREx1}
 \frac{1}{r} =s(u,E,L)= \frac{\alpha(E,L)}{u}.
\end{equation}
For the $E$ to be canceled in eq.(\ref{WeierVu2-eq}), then the following equation must be satisfied
\begin{equation}
\label{Weierc}
 c(E,L) =-\frac{Em\alpha(E,L)}{L^2}=\frac{g_3(E,L)}{2},
\end{equation}
or
\begin{equation}
\label{Weierg3}
 g_3(E,L) =-\frac{2Em\alpha(E,L)}{L^2}.
\end{equation}
Accordingly, the potential of eq.(\ref{WeierVu2-eq}) for this case is
\begin{equation}
\label{WeierVEx1}
  V(r,E,L)=-\frac{L^2}{2mr^2}+\frac{g_2(E,L)L^2}{2mr\alpha(E,L)}-\frac{2L^2\alpha(E,L)}{mr^3}.
\end{equation}
As a special case, consider the case with
\begin{equation}
\label{WeierCase1}
    g_2(E,L)=0, \hspace{10mm} \alpha(E,L)=\frac{m\kappa}{2L^2},
\end{equation}
where $\kappa$ is coupling constant. Then the potential in eq.(\ref{WeierVEx1}) can be written as
\begin{equation}
\label{WeierVExb1}
  V(r,L)=-\frac{L^2}{2mr^2}-\frac{\kappa}{r^3}.
\end{equation}
From eq.(\ref{WeierREx1}) and eq.(\ref{WeierCase1}), the expression of $r$ takes the following form
\begin{equation}
    r=\frac{m\kappa}{2L^2 \wp(\varphi;0,g_3)}
\end{equation}

For a closed orbits, $g_3(E,L)$ must take certain values, such that the Weierstrasse function is $2\pi n$ -periodic. This means that $g_3(E,L)=C^{(i)}_1,C^{(i)}_2,...$,
where $C^{(i)}_1$ is the value of $g_3$ such that the  Weierstrasse function is $2\pi $-periodic, $C^{(i)}_2$ is the value of $g_3$ such that the
Weierstrasse function is $4\pi $-periodic, and so on. The values of $i=1,2,..$ starts from $i=1$ where $g_3$ has the minimum value. There are infinite number of possible values
for $g_3$ correspond to each $2\pi n$-periodicity. This means according to eq.(\ref{Weierg3}), that the orbit is only closed orbit when the energy is related to the angular
momentum by the following relation
\begin{equation}
\label{WeiercEg3}
E =-\frac{2L^6C^{(i)}_n}{m^3 \kappa^2}, \hspace{10mm} i,n=1,2,3,...
\end{equation}
The above equation, together with the previous argument mean that; there are infinite number of possible energy values correspond to any value of the angular momentum
which give closed orbits, however, the energy spectrum is discrete, and it is determined by the values of $C^{(i)}_n$. The same results can be reached when solving
directly the Newton's orbit equation for the case of a force derived from the potential in eq.(\ref{WeierVExb1}). This solution  serves as a consistency check.
\begin{figure}[tbh]
\begin{center}
\includegraphics[bb=280 28 400 400,scale=0.65]{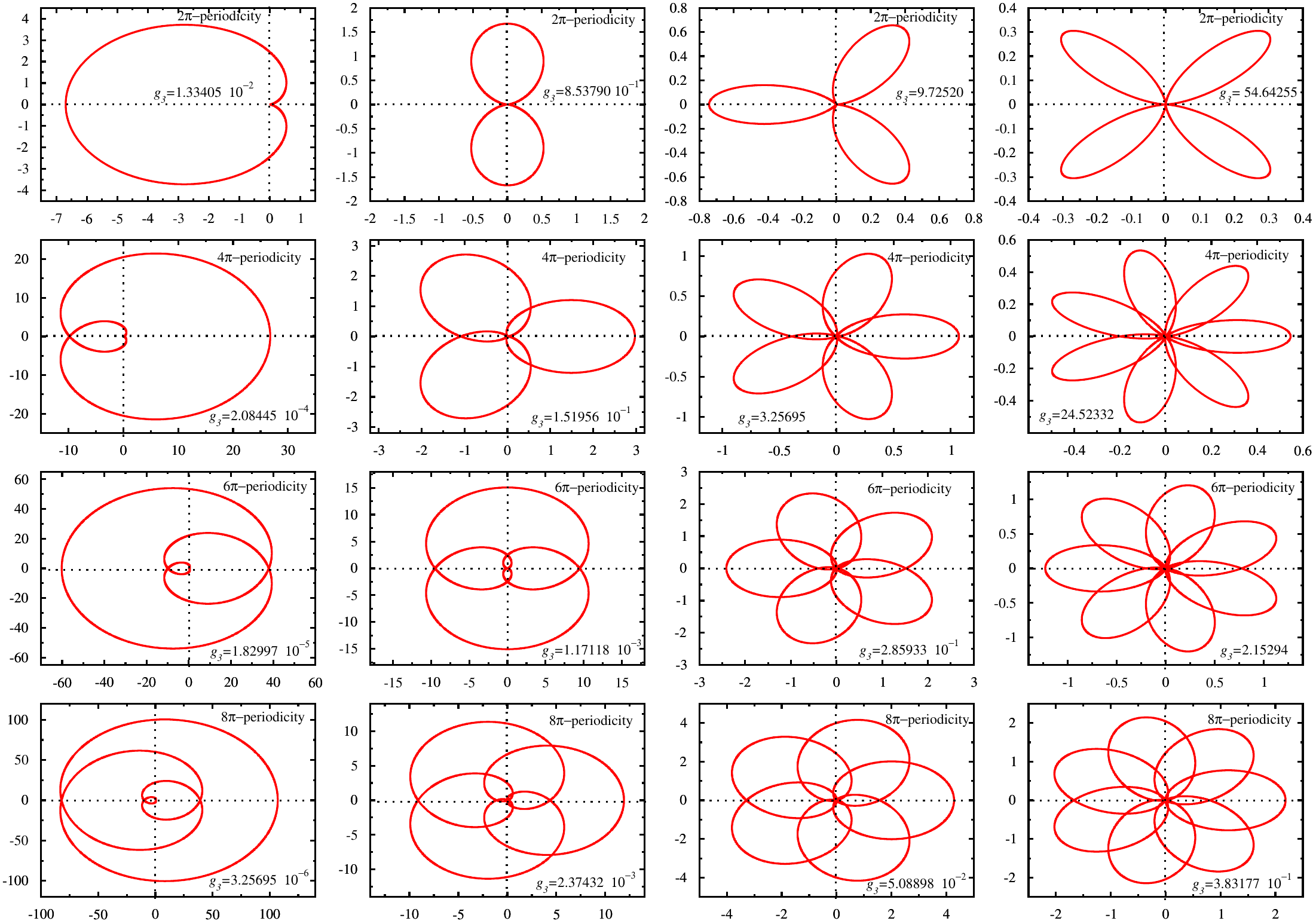}
\vspace{5mm}
\caption{\it Closed orbits for a particle moving under the action of the potential $ V(r,L)$ of eq.(\ref{WeierVExb1}), when $g_2=0$.
The first upper row is for $2\pi$-periodicity Weierstrasse function. The first graph from the left is for the lowest possible value of $g_3$,
the second graph at the same row is for the second lowest value of $g_3$, and so forth. The second,
third and fourth rows are for $4\pi$, $6\pi$ and $8\pi$-periodicity respectively.
In all the graphs we put $\alpha(E,L)=m\kappa/2L^2=1$.}
\label{fOrbWeierFirst}
\end{center}
\end{figure}
Figures \ref{fOrbWeierFirst} shows that the orbits are closed indeed for certain values of $g_3$.

Another special case when
\begin{equation}
g_2=\frac{4\lambda^4}{3}, \hspace{10mm} g_3=\frac{8\lambda^6}{27}, \hspace{10mm} \alpha(E,L)=\frac{m\kappa}{2L^2},
\end{equation}
where $\lambda$ any rational number. Accordingly, the potential of eq.(\ref{WeierVEx1}) can be written as
\begin{equation}
\label{WeierVEx2}
  V(r,L)=-\frac{L^2}{2mr^2}+\frac{4L^4\lambda^4}{3m^2\kappa r}-\frac{\kappa}{r^3}.
\end{equation}
For this, using eq.(\ref{WeierREx1}),eq.(\ref{WeierEx1}), and eq.(\ref{WeierHo}) give
\begin{equation}
\label{rWeierEx2}
  r=\frac{m\kappa}{2L^2\wp(\varphi;\frac{4\lambda^4}{3},\frac{8\lambda^6}{27})}=\frac{3m\kappa}{4L^2\lambda^2\left(1+\frac{3}{2}\cot(\lambda\varphi)^2\right)}.
\end{equation}

\begin{figure}[tbh]
\begin{center}
\includegraphics[bb=280 28 400 400,scale=0.6]{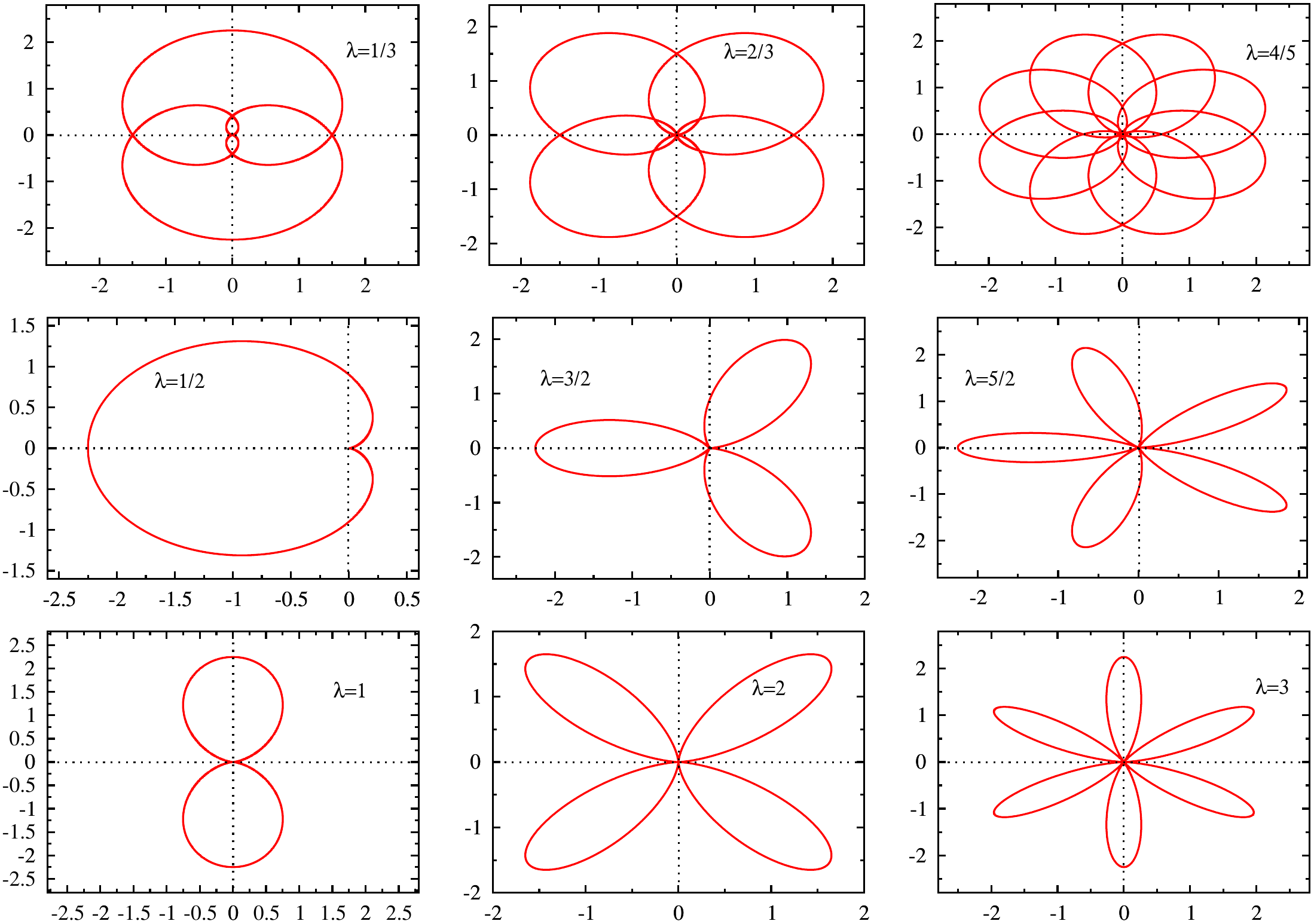}
\vspace{5mm}
\caption{\it Closed orbits for a particle moving under the action of the potential $ V(r,L,\lambda)$ of eq.(\ref{WeierVExb1}) for different values of $\lambda$ when
$g_2=4\lambda^4$. The first upper row is for $2\pi$-periodicity Weierstrasse function. The first graph from the left is for the lowest possible value of $g_3$,
the second graph at the same row is for the second lowest value of $g_3$, and so forth. The second, third and fourth rows are
for $4\pi$, $6\pi$ and $8\pi$-periodicity respectively. In all the graphs we put $\alpha(E,L)=m\kappa/2L^2=1$.}
\label{fOrbWeierSecond}
\end{center}
\end{figure}
The energy spectrum for this case is
\begin{equation}
\label{WeiercEg3g2}
E =-\frac{16L^6\lambda^6}{27m^3 \kappa^2},
\end{equation}
Again here, there are infinite number of possible energy values correspond to any value of the angular momentum which give closed orbits, however,
the energy spectrum is discrete, and it is determined by the value of $\lambda$, which is a rational number.
\section{Possible applications in old quantum mechanics and astrophysics}
In this section, the alternative orbit equation will be used in examples in Bohr Sommerfeld quantization, and stellar kinematics. However, the aim here is to explore the usefulness of the present treatment in future investigations in quantum mechanics, and astrophysics. 
\subsection{Closed orbits and Bohr Sommerfeld quantization }
In non-relativistic as well as relativistic cases, the Bohr Sommerfeld quantization has very few applications, mainly because of the limited number of closed orbits,
even when we add cases like the motion of a particle on a cone with a Kepler potential, or an isotropic harmonic potential at the tip of the cone \cite{Con}.
It has been demonstrated so far that there are so many potentials with nice features, and lead to closed orbits, which has been highlighted by various figures.
Therefore, the number of cases under which the Bohr Sommerfeld quantization is applicable can be hugely increase, provided that the integrals in eq.(\ref{Bohr-quan2})
are doable. For angular momentum, the quantization is straightforward, which is
\begin{equation}
\oint d\varphi p_\varphi=L\int_{0}^{2\pi}d\varphi=2\pi L=l h,
\end{equation}
which gives
\begin{equation}
\label{LBohr}
    L=\hbar l
\end{equation}
\begin{eqnarray}
\label{SommerPr}
\oint p_r dr&=&\int_0^{2\pi n}p_r\frac{dr}{d\varphi}d\varphi=\int_0^{2\pi n}p_r\frac{dg}{d\varphi}d\varphi \nonumber\\&=& \oint p_r\frac{dg(u,E,L)}{du}du.
\end{eqnarray}
As it is well known
\begin{equation}
p_r=\pm\sqrt{2Em-\frac{L^2}{r^2}-V(r)},
\end{equation}
therefore, from eq.(\ref{SommerPr}) one obtains that
\begin{equation}
\label{SommerPr2}
  \oint du p_r\frac{dg(u,E,L)}{du}=\pm\oint du g'(u,E,L)\sqrt{2Em-\frac{L^2}{g(u,E,L)^2}-V(u,E,L)}.
\end{equation}
An important observation can be made from the previous orbit figures for LSODE-periodicity orbits; when $\lambda$ is an integer, orbits have a certain numbers of sectors.
In addition, it takes $\Delta \varphi=\pi/\lambda$ for a particle to move from the point where $r=r_{min}$ to $r=r_{max}$. This can be verified from the fact that
$r=r(\cos(\lambda \varphi)$. The value of $r=r_{min}$ to $r=r_{max}$ can be obtained by applying the following condition
\begin{equation}
    \frac{dr}{d\varphi}= \frac{dg(u,E,L)}{d\varphi}=\frac{dg(u,E,L)}{du}\frac{du}{d\varphi}=0,
\end{equation}
Accordingly, for optimum $r$, either
\begin{equation}
\label{SommerOpt1}
  \frac{dg(u,E,L)}{du}=0,
\end{equation}
or
\begin{equation}
\label{SommerOpt2}
{du}\frac{du}{d\varphi}=0,
\end{equation}
or both. The previous discussion leads to
\begin{equation}
\label{SommerPr3}
    \oint p_r\frac{dg(u,E,L)}{du}du=\pm 2\lambda \int_{u_{min}}^{u_{max}}du g'(u,E,L)\sqrt{2m(E-V(u,E,L))-\frac{L^2}{g(u,E,L)^2}}.
\end{equation}
Using eq.(\ref{Vu-eq}), the above equation can be written as
\begin{eqnarray}
\label{SommerPr4}
 \oint p_r\frac{dg(u,E,L)}{du}du&=& \pm 2\lambda\int_{u_{min}}^{u_{max}}du\frac{L g'(u,E,L)^2}{g(u,E,L)^2} \sqrt{2ua(E,L)-2c(E,L)-\lambda^2u^2}\nonumber\\&=&n_r h.
\end{eqnarray}
To check the validity of this treatment, consider the case of the Kepler potential, then $g(u,E,L)=1/u$, $\lambda=1$, $a(E,L)=m\kappa/L^2$, and $c(E,L)=-Em/L^2$.
The values of $u_{min}=a(E,L)-b(E,L)$, and $u_{max}=a(E,L)+b(E,L)$, where $b(E,L)$ is given by eq.(\ref{KepA}). Using the table on integrals \cite{Gradshteyn},
the integral in eq.(\ref{SommerPr4}) can be obtained,  then solving for $E$ gives
\begin{equation}
\label{KeperBohr}
    E=-\frac{m\kappa^2}{2\hbar^2 (n_r+l)^2}=-\frac{m\kappa^2}{2\hbar^2 n^2}
\end{equation}

The simplest example of the quantization of an energy angular momentum dependent potential system is the potential in eq.(\ref{Vr2VEL}). For this case, the quantization can simply
be preformed by replacing $\kappa$ by $\kappa(E,L)$, then eq.(\ref{KeperBohr}) can be written as.
\begin{equation}
\label{KeperBohr2}
    E=-\frac{m\kappa(E,L)^2}{2\hbar^2(n_r+l)^2}=-\frac{m\kappa(E,L)^2}{2\hbar^2 n^2}.
\end{equation}
If $\kappa(E,L)$ assumed to have the following form
\begin{equation}
\label{KeperEL}
  \kappa(E,L) =\kappa_0\frac{EL^{2}}{m \kappa_0^2},
\end{equation}
where $\kappa_0$ is the force constant, which is simply in this case $\kappa_0=Ze^2/4\pi\epsilon_0$, and $EL^{2}/(m \kappa_0^2)$ is a unitless quantity.
Accordingly, eq.(\ref{KeperBohr2}) and eq.(\ref{KeperEL}) give
\begin{equation}
\label{KeperBohr3}
    E=-\frac{m\hbar^2 \kappa_0^2 n^2}{l^4}.
\end{equation}
What is interesting about the energy spectrum for this system is; it proportional to $n^2/l^4$, which is a rational number.
Moreover, the energy has infinite degeneracy, because there is infinite values of $n$ and $l$, such that
$n^2/l^4$ equal to a certain rational number.
\subsection{Modified Newtonian gravity, and possible applications in astrophysics}
Assuming that the gravitational potential is modified to an energy dependent
potential, like the one in eq.(3.36). There are infinite possibilities for such modifications. However, we are choosing an example without much scrutiny in relation to observation. let us assume that the Newtonian gravity is modified to
\begin{equation}
\label{GalaxyVmod}
   V(r,E,L)= -\frac{k_0 \exp\left(\frac{EL^2}{2M\kappa_0^2}\right)}{r},
\end{equation}
where $M$ is the mass of the source, $m$ is the reduced mass, and $k_0=G_0 m M$. To make the example even more simple, only circular motion is considered. Then, for this case,
the Newton's second law gives
\begin{equation}
\label{vcVmod}
   v_c^2 \exp\left(\frac{ m r_c^2 v_c^4}{4 G_0^2 M^3}\right)-\frac{G_0M}{r_c}=0,
\end{equation}
where $r_c$ is the radius of the circular orbit around the center of the disk with velocity $v_c$, and $A$ is unitless constant that can fixed from a
possible observation. The above equation can be solved for $v_c$, the result is
\begin{equation}
\label{vcVmodSol}
  v_c=\sqrt{\frac{G_0}{r_c}}\sqrt[4]{\frac{M^3W[m/2M]}{m}}
\end{equation}
where $W$ is the Lambert-W function which is also named as log product function. When comparing the result velocity in eq.(\ref{vcVmodSol}) with the one that can be obtained from
Newtonian gravity, which is $v_c=\sqrt{G_0M/r_c}$, we find that the difference between the two solution is very small when $m\ll M$. This can be shown by expanding
the Lambert-W function. Then eq.(\ref{vcVmodSol}) can be written as
\begin{equation}
\label{vcVmodSol2}
v_c=\sqrt{\frac{G_0 M}{r_c}}\left(1-\frac{m}{8M}\right)+O\left(\frac{m^{2}}{M^{2}}\right).
\end{equation}
In the above equation, it is obvious that the second leading term is small, therefore one would ask what is the importance of such example. The answer is, the second term depends on $m$, which violates the Einstein's equivalence principal. When the two bodies in the system haves equal masses, then $m=M/2$,  which gives the maximum deviation from the Newtonian gravity, however it is still very small. Moreover, such deviation is not supported by observation.

One of the most intriguing phenomena in modern physics is stellar kinematics in galaxies. According to the observation, stars are moving faster than predicted by calculations based on the mass of the visible material \cite{Z1}. To explain mass deficit, the hypothesis of dark matter was proposed \cite{darkCon}. An alternative approach called "modified newtonian dynamics", or shortly as MOND, which is based on modifying the Newton's laws to give an account for the stellar kinematics \cite{Milgrom}.

At this point, it is far from clear if modifying the Newtonian gravitational potential would give a successful description of the stellar kinematics, especially if it is
modified to an energy and angular momentum dependent potential. The author would like to stress here, that the following example is by no means aiming at
presenting any alternative, or modification to the gravitational potential $V(r)=-\kappa_0/r$. It is aiming at demonstrating that a modification of a potential to an
energy and angular momentum dependent potential, can lead to velocity distant curve that could be explained wrongfully as a sign of a missing mass, or a mass that is not accounted for.

Consider a very thin disk $z$ constitute of particles that are not interacting with their neighbors, but moving under the action of the collective gravitational field of all the
particles in the disk. If the disk rotates in a constant angular velocity $\omega_c=v_c/r_c$, then the equation of motion for a test particle with mass $m$, and circular orbit is
\begin{equation}
\label{GalaxyDisk1}
    \frac{\kappa}{r_c^2} =\frac{\tilde{G_0}M(r_c) m}{r_c}=\frac{m v_c^2}{r_c}=m,
\end{equation}
where $G_0$ is the gravitation constant.
For $\omega_c=constant$ at any point on the rotating disk, the density of the disk has to obey the following relation
\begin{equation}
\label{GalaxyDiskDensity}
    \rho(r)=\rho_0 \frac{r}{R}
\end{equation}
where $\rho_0$ is a constant, and $R$ is the radius of the disk. At textbook exercise can show that the outer mass of the disk has no contribution to the
equation of motion of the test particle. The interior mass at $r=r_c$ is given by the following relation
\begin{equation}
\label{GalaxyEqM1}
 M(r_c)=\int_{0}^{r_c} 2\pi r^2\frac{\rho_0}{R}z=2\pi z \frac{r_c^3}{3R}\rho_0.
\end{equation}
Accordingly, the equation of motion for a particle with a circular orbit of a radius $r_c$ is
\begin{equation}
\label{GalaxyEqM1}
\frac{m v_c^2}{r_c}=\pi\rho_0\frac{ r_c^3}{3} z\frac{G_0 m}{r_c^2},
\end{equation}
which leads to
\begin{equation}
\label{Galaxyvc1}
 v_c=r_c\sqrt{\frac{2\pi z G_0 \rho_0}{3R}}=r_c \omega_c.
\end{equation}

Now assuming that the gravitational potential is modified to an energy dependent potential, like the one in eq.(\ref{GalaxyVmod}). For a circular orbits, the constants of motion are given by the following equations
\begin{equation}
\label{LcEc}
  E=\frac{m v_c^2}{2}, \hspace{10mm} L= m v_c r_c.
\end{equation}
It is clear that the expression $EL^2/(M\kappa_0^2)$ in eq.(\ref{GalaxyVmod}) is unitless. For a particle in a thin disk problem,
the modified gravitational potential leads to the following expression for the velocity
\begin{equation}
\label{Galaxyvemod}
    v_c=\sqrt[4]{\frac{8\pi^3 G_0^2 r_c^7z^3\rho_0^3 W\left[\frac{3mR}{2\pi r_c^3 z\rho_0}\right]}{27 m R^3}},
\end{equation}
By plotting $v_c$ versus $r_c$ from  eq.(\ref{Galaxyvemod}), and $v_c$ versus $r_c$  for the Newtonian gravity, it can be realized that near the center of the disk, the two curves differ. An experimentalist who is not informed about any modification in the gravity law, would read the two curves as a difference in density,
where the matter is less dense near the center for the modified case, although "in fact", the density is equal for the two cases.
\\
\begin{figure}[h]
\begin{center}
\includegraphics[bb=300 12 400 400,scale=0.6]{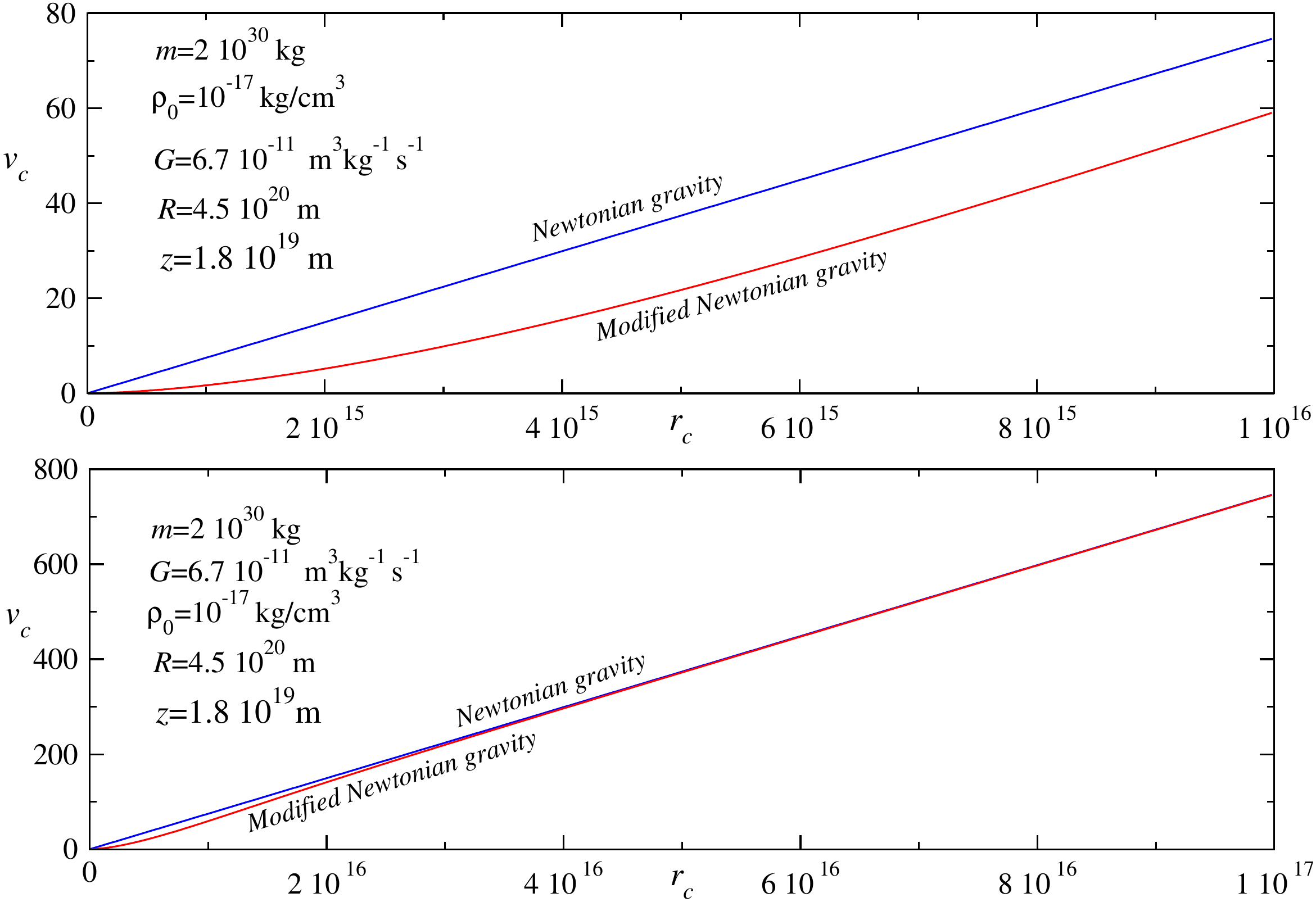}
\vspace{5mm}
\caption{\it The graphs for $v_c$ velocity (m/sec units) versus distant $r_c$ (m units) for non-interacting bodies on a galactic thin disk with a radius $R$, and thickness $z$ .
The test mass is equal to the mass of the sun. The upper panel shows the curves for Newtonian gravity and the modified Newtonian gravity near the center of the disk, in a
distant range less than 1 pc. The lower panel show the same curves in distant range of 10 pc.}
\label{fOGalaxy}
\end{center}
\end{figure}

The values $z$ , $R$, $\rho_0$ in the figure \ref{fOGalaxy} are chosen to be roughly close to the measured values for the Milky way's galactic disk, while $m$ is equal to the
mass of the sun in kg.Indeed the result is opposite to the observation. It shows in fact that this modification does not offer any explanation for the galactic kinematics. In addition, the modification cause a violation to the Einstein equivalence principal. The only justification for this argument is a mere  demonstration that a modification to
the Newtonian gravity law could lead to the wrong conclusion about an extra mass, or a mass a deficit.
\section{Summary and Conclusions}
By tracing the concept of potential function from the early beginning to the present time, one can conclude that it is a concept that was brilliantly invented to explain observation,
especially in relation to closed orbits. The Bertrand theorem concluded that;  the Kepler potential, and the isotropic harmonic oscillator potential are the only systems under which
all the orbits are closed. It was never stressed enough in the physical or mathematical literature that this is only true when the potential is independent of the initial conditions
of motion, which, as we know, determine the values of the constants of motion $E$ and $L$. In other words, the Bertrand  theorem is correct only when $V\equiv V(r)\neq V(r,E,L)$. In fact it has been proved in this work that there are infinitely many energy angular momentum potentials $V(r,E,L)$ that lead to closed orbits. Reaching to this conclusion was done by generalizing the well known substitution $r=1/u$ in Newton's orbit equation to a general substitution $r=g(u,E,L)$ or $r=1/s(u,E,L)$ in the equation of motion. This led to the derivation of what is equivalence to Newton's orbit equation, which we called the "alternative orbit equation". It is not written in terms  second order differential equation, but in terms of an energy angular momentum dependent potential $V(u,E,L)$ (see eq.(\ref{Vu-eq}), and eq.(\ref{Vu2-eq})). If $u$ is a periodic function $\varphi$,
such that it obeys a second order differential equation of the form $u^{\prime\prime}=\gamma(u,E,L)$, where $\gamma(u,E,L)$ depends on the periodicity of the orbit, then the orbits are closed, because  $r(\varphi)=r(\varphi+2\pi n)$.
In the literature, studying different systems using Newton's orbit equation is done by using the force or potential as an input. What is different in this work is using the periodicity of the orbit given by $\gamma(u,E,L)$, and the choice of $r=g(u,E,L)$ or $r=1/s(u,E,L)$ as inputs, and then find what potentials $V(u,E,L)=V(r,E,L)$ that give closed orbits. Under such consideration, for each periodicity characterization function given $\Omega(u,E,L)=\int_a^u\gamma(\zeta,E,L)d\zeta$, there are infinite number of potentials $V(r,E,L)$ for a certain periodicity characterization function $\Omega(u,E,L)$ that give closed orbits. Having said that, this does not mean that the choice of $r=1/s(u,E,L)$ is absolutely random, because $r(\varphi)$ must be real and positive for any value of $\varphi$. This condition is a constant reminder that such problems must be handled with care, because the condition restricts the domain of the energy $E$ for a given angular momentum $L$.

The most important finding of this work is the expression of the energy angular momentum $V(u,E,L)$ in eq.(\ref{Vu-eq}), or an alterative form in eq.(\ref{Vu2-eq}).
One of them is more suitable for certain application, and the other one is more suitable for others. However, both of them are called here the alternative orbit equation. In this work, the expression of $V(u,E,L)$ in eq.(\ref{Vu2-eq}) is more practical to use in most of the chapters. The first application of the theorem for a certain $\Omega(u,E,L)$ is the for the case of a linear second differential equation orbits,  or the case
when $u=a(E,L)\lambda^{-2}+b(E,L)\cos\lambda \varphi$, where $\lambda$ is a rational number, then $\Omega(u,E,L)=a(E,L)u-c(E,L)-u^2\lambda^2/2$.
For such case, the best test for the theorem is show that it does not contradict with the Bertrand theorem when $V(u,E,L)=V(u)$. Indeed this is the case when $s(u,E,L)=u^n$,
where the alternative orbit equatio can only be satisfied in two cases. The first case when
$n=1$ and $\lambda=1$, which leads to the Kepler potential $V(u)=V(r)=-\kappa/r$. The second case when $n=1/2$ and $\lambda=2$, which leads to the isotropic harmonic potential
$V(u)=V(r)=m\omega^2r^2/2$. Moreover, for each case, the values of $a(E,L)$ and $b(E,L)$ can be obtained from eq.(\ref{Vu2-eq}), they are exactly the same expressions that can be
obtained from solving directly the Newton's orbit equation. 

For orbits of such periodicity, a special case solution of the form  $s(u,E,L)=\tilde{s}(u,E)/L$ was considered, then the alternative orbit equation gives the potential of the form $V(u,E)=V(rL^{-1},E)$, which is called the rescaling potential. Two examples of such potentials were studied carefully. For the first example $V(u)\sim u^2$, an interesting figures for orbits where obtained, like figures \ref{fOrb05}, \ref{fOrbAll}, moreover, it has been shown that for real positive $r$, there is a restriction on the value of the energy $E$ (see eq.(\ref{ECon1})). For the lower possible value of $E$, the orbits have a certain point, or points for which $r\rightarrow \infty$, this is shown by figure \ref{fOrbOpen}. As for the second example, the solution is more complicated as $V(rL^{-1},E)$ contains the term $\arccsch(cr)$ (see eq.(\ref{VrSinh1})). For such potential, solving the Newton's orbit equation to get the orbit graphs is a very difficult task, or maybe not possible. However, $r(\varphi)$ is an input in our treatment, therefore plotting the orbits is possible, as one cane see in figures \ref{fOrbSinhEq1},\ref{fOrbSinhEq2}, which is a proof of the usefulness of this approach.

For the same $\Omega(u,E,L)$, a more general solution when $s(u,E,L)$ is not factored. A simple example in this context is $s(u,E,L)=u/(\alpha_0(E,L)+\alpha_1(E,L) u)$ leads to
a potential of the form $V(r,E,L)=\sum_{n=1}^{4}A_n(E,L) r^{-n}$, where $A_n(E,L)$ are constants given by eq.(\ref{An}). The possibility of reducing this potential to a two
term-potential was investigated. It has been proved that this is not always possible. The only physical case are three cases, the first case when $A_2=A_4=0$, then the orbits are
shown by figure \ref{fOrbEmpFirst}, the second case when $A_1=A_4=0$, then the orbits are shown by figure \ref{fOrbEmpSecond}, the third case when $A_1=A_3=0$, then the orbits are
circles. For one-term potential, it was proved that this is only possible if $\alpha_1(E,L)=0$, then the potential is for  what we call a modified Kepler problem,
with $\kappa\rightarrow \kappa(E,L)$, which is also leads to a closed orbit for arbitrary $\kappa(E,L)$.

The second application of the theorem is on the Weierstrasse periodicity orbits with $\Omega(u,E,L)=2 u^3-g_2(E,L)u 2^{-1}+g_3 2^{-2}$. For this application, one simple
examples where studied, which is for $s(u,E,L=\alpha(E,L)/u$, then the potential is given by eq.(\ref{WeierVEx1}). Case in this example is when $g_2=0$, then the value of $g_3$ that closed the orbit are given by figure \ref{fOrbWeierFirst}. This leads to an interesting result, which is a discrete energy spectrum $ E=-2L^6C^{(i)}_n m^{-3} \kappa^{-2}$  as a condition for the orbit to be closed. Another important issue here is the potential for this case $V(r,E,L)=-(L^2/2mr^2)-\kappa/r^3$, is a system that could be solved directly using the Newton's orbit equation. The result is exactly the same as the one the we got using the alternative orbit equation. This is an important checking for the validity of our approach. The second case for this example is when $g_2=4\lambda^4/3$, and $g_3=8\lambda^6/27$. For this case, the energy is $ E =-16L^6\lambda^6/27m^3 \kappa^2$, which is also discrete. The orbits for the second case are shown by figure \ref{fOrbWeierSecond}

One of the important motivation to for this work is to increase the number of systems that could be quantized using the Bohr Sommerfeld quantization. That is because this quantization is applicable when the orbit is closed. The quantization of angular momentum is straightforward  $\oint d\varphi p_\varphi=2\pi L=l h$. Using the first form of the theorem in
eq.(\ref{Vu-eq}), the integral $\oint p_r dr=n_r h$ can be written in terms of the variable $u$. This enable us to quantized the system using $V(u,E,L)$ directly, and without the
involvement of $r$ (see eq.(\ref{SommerPr4}). The validity of the treatment was examined by calculating the energy from eq.(\ref{SommerPr4}) for the Kepler potential, then $g(u,E,L)=1/u$.
After preforming the integrating, we get the correct energy spectrum for the hydrogen atom $E=-m\kappa^2/2\hbar^2 n^2$. A new example also discussed, where $\kappa(E,L)$, replaced by
$\kappa_0(EL^2/m \kappa_0^2)$ in the Kepler potential, then the energy spectrum for this case can be simply obtained by replacing $\kappa_0\rightarrow \kappa(E,L)$, to get
$ E=-m\hbar^2 \kappa_0^2 n^2/l^4$.

The treatment at in 3.1 has shown that the Kepler potential can give a closed orbit by replacing the constant $\kappa_0$ with an energy angular momentum dependent constant $\kappa(E,L)$
can still lead to a closed orbit. One can argue that this can be done directly without using the theorem in eq.(\ref{Vu2-eq}), but by a mere replacement. One can argue also that such
replacement is a trivial matter, it is only prohibited by terms like being "a wild assumption, unusual, unacceptable, or does not comply with observation". The aim of this article is not to comply with what is "acceptable", the main aim here is to expand our kit of tools to describe
different phenomenon, for example it is true that the modification of the Kepler problem that was discussed in the previous paragraph is poorly motivated, however
it is a point of research that could have future applications yet to be discovered, especially when the solution of the Schr\"{o}dinger equation is impossible or hard to get, then using the treatment in 5.1 would be useful, this is left to a future work.

Observation still a loose expression, it is true that the discovery of Newton for the gravitational potential proved to be one of the biggest successes in theoretical physics, as it gives a correct account for the motion of celestial bodies, and beyond. One of the question is, how valid this potential is when we are speaking about distances on galactic scale, or at very short distances, shorter than any present experiments can examine. The other question is; can we modify the Newtonian gravity such that the observed stellar kinematics can be explained without the assumption of dark energy or dark matter, and without using the modified newtonian dynamics (MOND). Partly motivated by these issues, we discussed in 5.2  the possibility of modifying the Kepler problem. A modified  gravitational potential $V(r,E,L)=-k_0 \exp(EL^2/Mk_0^2)/r$, where $k_0=MmG$, and $EL^2/Mk_0^2$ is unitless, can still give closed orbits. The modified newtonian gravitation law was used on a thin disk that rotate with constant angular velocity, where each body moves under the action of the collective gravitational force produced by all the bodies constituting the disk, such that the force between neighboring bodies is negligible. The variables of the disk were taken roughly close to the variables of the galactic disk for the milky way (density, radius,..etc ). Moreover, it was assumed that the mass of each body constituting the disk is equal the mass of the sun. Accordingly, the tangential velocity $v_c$ was calculated as function of the distant from the center of the disk $r_c$. The results of the calculations are shown in figure \ref{fOGalaxy}, where the deviation from the Newtonian gravity is more noticeable near the center of the disk, with the non-Newtonian gravity gives smaller velocity than Newtonian gravity for any value of $r_c$. The difference between Newtonian and non-Newtonian gravity diminishes as $r_c$ increase. This means that our particular modification to the Newtonian gravity is a failure, because it cannot be supported by the present observation. In fact, the observed stellar kinematics indicates that the velocity $v_c$ increase with $r_c$ in such manner that cannot be accounted by the observed matter, and therefore the assumption of  dark matter deemed to be necessary. Then one would ask, what is the point of discussing such
modification. The answer is; if an experimentalist is not informed about any modification of the Newtonian gravity, then he would interpreted figure \ref{fOGalaxy} as a less
density than the one actually used  $ \rho(r)=\rho_0 r/R$  (see eq.(\ref{GalaxyDiskDensity})). Therefore, we can rule out modifying the Newtonian gravity to explain the observed stellar kinematics, although such modification is not available at the present time.  

Another possible area of future investigation is studying Bertrand potentials perturbed by a small additional potential which depends on the energy and angular momentum, such that the orbits stays closed. This is again can be done using the alternative orbit equation. It is expected that the results have an important practical applications.


\begin{thebibliography}{20}
\bibitem{Grant}
Grant, Edward, "Celestial Orbs in the Latin Middle Ages," Isis, 78(1987): 153–73; reprinted in Michael H. Shank, ed.,
The Scientific Enterprise in Antiquity and the Middle Ages, Chicago: Univ. of Chicago Pr., 2000. ISBN 0-226-74951-7.
\bibitem{Holton}
S.\ G.\ Brush, G.\ Holton, Physics, The Human Adventure: From Copernicus to Einstein and Beyond, third eddition, Rutgers University Press, 2001.
\bibitem{Gold}
H.\ Goldstein, Classical Mechanics, Second Edition, Addison Wesley Publishing Company, Inc., 1980.
\bibitem{Jammer}
M.\ Jammer, Concepts of Force, Courier Corporation, 1999.
\bibitem{Bertrand}
J.\ Bertrand, C.\ R.\ Acad.\ Sci. 77 (1873) 849.
\bibitem{Whittaker}
 E.\ T.\ Whittaker, A Treatise on the Analytical Dynamics of Particles and Rigid Bodies, fourth edition, Dover, New York, 1944.
 \bibitem{MAHOMED1}
F.\ M.\ Mahomed, and F.\ Vawad, Nonlinear Dynamics 21 (2000) 307.
\bibitem{Adler}
 R.\ Adler, M.\ Bazin, and M.\ Schiffer, Introduction to General Relativity, Second eddition, McGraw-Hill, 1975.
\bibitem{Jetzer}
 P.\ Jetzer, General Relativity, Lecture notes by Philippe Jetzer, 2011.
\bibitem{Sakurai}
J.\ J.\ Sakurai, Advanced Quantum Mechanics, First eddition, Addison-Wesley, 1967.
\bibitem{Schiff}
L.\ Schiff, Quantum Mechanics, Third edition, McGraw Hill. Inc., 1968.
\bibitem{Lawden}
D.\ F.\ Lawden, Elliptic Functions and Applications, Springer-Verlag New York Inc., 1989.
\bibitem{Akhiezer}
N.\ I.\ Akhiezer, Elements of the Theory of Elliptic Functions, American Mathematical Society.Providence.Rhode Island,
Transelations of Mathematical Monographs Vol.79, 1990.
\bibitem{Hancock}
H.\ Hancock, Lectures on the Theory of Elliptic Functions, First eddition, John Wiley \& Sons, 1910.
\bibitem{Wolfram}
Functions.wolfram.com, Introductions to WeierstrassInvariants.
\bibitem{Con}
 M.\ H.\ Al-Hashimi, and U.\-J.\ Wiese, Ann.\ Phys.\ 323 (2008) 82.
\bibitem{Gradshteyn}
 I.\ S.\ Gradshteyn, and I.\ M.\ Ryzhik, Table of Integrals, Series, and Products,
Fifth Edition, Academic Press 1996.
\bibitem{Z1}
E.\  Zwicky, ApJ 86 (1937) 217.
\bibitem{darkCon}
J.\ de Swart, G.\ Bertone, and J.\ van Dongen, Nature Astronomy 1 (2017) 0059.
\bibitem{Milgrom} 	
M.\ Milgrom, ApJ 270 (1983) 365.
\end{thebibliography}
\end{document}